\documentclass[iicol,sn-mathphys-num]{sn-jnl}

\usepackage{graphicx}
\usepackage{multirow}%
\usepackage{amsmath,amssymb,amsfonts}%
\usepackage{amsthm}%
\usepackage{mathrsfs}%
\usepackage[title]{appendix}%
\usepackage{xcolor}%
\usepackage{textcomp}%
\usepackage{manyfoot}%
\usepackage{booktabs}%
\usepackage{algorithm}%
\usepackage{graphics}
\usepackage{algorithmicx}%
\usepackage{algpseudocode}%
\usepackage{listings}%
\usepackage{longtable}

\newcommand\blfootnote[1]{%
  \begingroup
  \renewcommand\thefootnote{}%
  \footnotetext{#1}%
  \addtocounter{footnote}{0}%
  \endgroup
}

\begin{document}

\title{Trust and Transparency in AI: Industry Voices on Data, Ethics, and Compliance}

\author*[1]{\fnm{Louise} \sur{McCormack}}\email{louise.mccormack@adaptcentre.ie}
\author[2]{\fnm{Diletta} \sur{Huyskes}}\email{diletta.huyskes@unimi.it}
\author[3]{\fnm{Dave} \sur{Lewis}}\email{dave.lewis@adaptcentre.ie}
\author[1]{\fnm{Malika} \sur{Bendechache}}\email{Malika.bendechache@adaptcentre.ie}
\affil[1]{\orgdiv{ADAPT Research Centre}, \orgname{School of Computer Science, University of Galway}, \orgaddress{\city{Galway}, \country{Ireland}}}
\affil[2]{\orgname{University of Milan}, \orgaddress{\city{Milan}, \country{Italy}}}
\affil[3]{\orgdiv{ADAPT Research Centre}, \orgname{Trinity College Dublin}, \orgaddress{\city{Dublin}, \country{Ireland}}}
\abstract{The EU Artificial Intelligence (AI) Act directs businesses to assess their AI systems to ensure they are developed in a way that is human-centred and trustworthy. The rapid adoption of AI in the industry has outpaced ethical evaluation frameworks, leading to significant challenges in accountability, governance, data quality, human oversight, technological robustness, and environmental and societal impacts. Through structured interviews with fifteen industry professionals, paired with a literature review conducted on each of the key interview findings, this paper investigates practical approaches and challenges in the development and assessment of Trustworthy AI (TAI). The findings from participants in our study, and the subsequent literature reviews, reveal complications in risk management, compliance and accountability, which are exacerbated by a lack of transparency, unclear regulatory requirements and a rushed implementation of AI. Participants reported concerns that technological robustness and safety could be compromised by model inaccuracies, security vulnerabilities, and an overreliance on AI without proper safeguards in place. Additionally, the negative environmental and societal impacts of AI, including high energy consumption, political radicalisation, loss of culture and reinforcement of social inequalities, are areas of concern. There is a pressing need not just for risk mitigation and TAI evaluation within AI systems but for a wider approach to developing an AI landscape that aligns with the social and cultural values of the countries adopting those technologies.
}

\maketitle

\blfootnote{\textbf{Note:} This work has been accepted for publication in \emph{AI and Society}.}

\section{Introduction}\label{sec1}

Artificial Intelligence (AI) is being increasingly adopted by a variety of industries, including essential sectors like healthcare, education, transport, and climate action, necessitating regulations to ensure the safe, ethical, and effective use of AI systems\cite{eu_ai_act_2024}.

In Europe, existing regulations covering these sectors are being supported by the introduction of the Artificial Intelligence Act (AI Act) (\cite{euaiact}) which are further enhanced by the General Product Safety Regulation (\cite{eugps}) and the Product Liability Directive (\cite{epld}), which both explicitly include aspects of software. The AI Act seeks to enable seamless access of AI products to the single market while protecting health, safety and fundamental rights, as well as the environment and democracy rights.

Preceding the introduction of the AI Act, recent years have seen intense activity in developing principles and guidelines that claim to support trustworthy AI systems, sometimes under the label of ethical or responsible AI (\cite{Correa2023}). Where adopted, these were typically voluntary measures, with little external validation and often appealing to subjective or contested forms of behaviour. This also led to accusations of ethics washing (\cite{schultz2024digital}). In contrast, on entering into law in August 2024, the EU’s AI Act(\cite{euaiact}) states that it is a  comprehensive legally binding set of rules for AI systems that integrates Europe’s existing legal protections of fundamental rights (based on the European Charter of Fundamental Rights (CFR)(\cite{Charter2012}) with its existing framework for harmonised product safety certification across the European single market. The requirement for AI products to address the protection of fundamental rights can be considered the practical transposition of many of the concerns addressed by prior Trustworthy AI initiatives into an enforceable legal framework. However, with the AI Act having entered into law in August 2024 and the enforcement of its different provisions being scheduled for phased introduction over a 36 month period\cite{eu_ai_act_2024}, it will be several years before concrete enforcement experience and case law becomes evident. 

Trustworthy AI (TAI) in this context therefore refers to an ethical, human-centred approach to artificial intelligence as outlined in the EU's Ethical Guidelines for Trustworthy AI (\cite{EUEGTAI}) published in 2019. These guidelines emphasise the importance of developing AI systems that are lawful, ethical, and robust. TAI focuses on ensuring that AI respects human rights, fosters transparency and accountability, and promotes fairness and inclusivity. By adhering to these principles, TAI aims to mitigate risks and enhance the positive impact of AI on society.

The seven principles of Trustworthy AI, as outlined by the EU's Ethical Guidelines for Trustworthy AI, are:

\begin{enumerate}
    \item Human Agency and Oversight
    \item Technical Robustness and Safety
    \item Privacy and Data Governance
    \item Transparency
    \item Diversity, Non-discrimination, and Fairness
\item Societal and Environmental Well-being 
\item Accountability
\end{enumerate}
There are several established methods (\cite{mccormack2024ethical}) and metrics (\cite{mccormack2024comprehensive}) to evaluate specific aspects of TAI. However, a significant gap remains in the availability of processes for a comprehensive evaluation of TAI. Existing methods often address isolated components, such as fairness, transparency, or accountability, but fail to provide an integrated framework that covers all ethical and technical dimensions of trustworthy AI systems. This lack of holistic evaluation frameworks presents challenges in ensuring that AI systems align with the full spectrum of TAI principles.

Additionally, TAI evaluation faces multiple challenges, including a lack of standardisation of metrics and methods, reliance on manual questionnaires, the need for use-case-specific evaluation methods, and fragmented AI development processes which cause issues relating to the accountability of evaluation (\cite{mccormack2024ethical}). Researchers have also found a lack of suitable industry tools, highlighting the impractical level of human effort required to make existing toolkits work to mitigate bias effectively (\cite{harris2023mitigating}), and the difficulties in translating real-world use cases into the quantifiable metrics required by these toolkits (\cite{deng2022exploring}). These challenges hinder the practical implementation of comprehensive TAI evaluation frameworks. AI systems are diverse, and there are many challenges associated with assessing them for trustworthiness. To address these challenges, researchers suggest a need for Standard Developing Organisations (SDOs) mandating ethical disclosures and ensuring minimum standards for testing, documentation, and public reporting (\cite{laux2024three}). One such standard is ISO/IEC 42001 (\cite{ISO42001}), which offers high-level guidance for compliance with the EU AI Act, in a similar way as ISO27001 guides compliance for information security management in line with the European General Data Protection Regulation(GDPR)\cite{lopes2019implementation}.To conduct empirical studies into industry preparedness for more strongly enforced trustworthy AI measures, we have conducted this study in line with the EU’s precursor framework for Trustworthy AI, which benefits from a degree of existing adoption, familiarity and implementation experience. This paper seeks to understand the challenges and trends that industry professionals face when evaluating AI systems for trustworthiness.

\section{Research Questions}\label{sec2}

To guide the investigation into the challenges and trends that industry professionals face when assessing AI systems for trustworthiness, this research aims to answer the following research questions:
 
(R1:) What are the primary challenges industries face regarding data acquisition, quality, preparation, and provenance in AI systems, and how do these challenges influence the trustworthiness of their AI-based solutions?

Data is a key part of AI systems and relates directly to its trustworthiness. Identifying data challenges in AI systems will help in developing strategies to enhance data management practices, thereby improving the trustworthiness of AI systems.
 
(R2:) How do industry professionals perceive the seven principles of Trustworthy AI?
This research question aims to understand how industry professionals perceive each of the seven principles, including identifying their priorities, challenges and comprehension of each.  

(R3:) What current practices and challenges exist in assessing compliance with AI-related standards and regulations (such as ISO27001 and GDPR), and how can organisations improve their processes to foster trustworthy AI systems?

Compliance with standards is essential for TAI, however regulating a fast-paced technology such as AI introduces new complications. To help inform the design of assessments for TAI, this question seeks to understand the current processes for evaluating areas such data security which is relevant to the security of AI systems also.

\section{Methodology}\label{sec3}
We invited industry professionals working in AI-related roles to participate in one-hour structured interviews. These professionals were selected through the research team’s existing professional networks to represent a variety of tech industries and job functions. We did not exclude any sectors in our selection, but did target sectors with high AI adoption and focusing on interviewing professionals working across a diversity of applications of AI. The goal of the interviews was to gain insights into the challenges and trends in assessing data use for Trustworthy AI (TAI). Conversations with the 15 professionals were audio recorded, with notes taken during the interviews.
 
To ensure compliance with ethics guidelines and GDPR regulations, the University of Galway’s Ethics Committee approved the outreach approach, which involved sending invitations via email and LinkedIn. Participants provided informed consent by signing consent forms prior to taking part in the research.
 
The insights from these 15 interviews were compiled by the research team and structured into key thematic sections. Empirical evidence suggests that the majority of themes (over 90\%) can typically be identified within the first twelve interviews\cite{guest2006many} with proposed methods to evaluate if saturation has been reached\cite{guest2020simple}. Further research emphasises that sample size decisions are inherently situated and cannot be predetermined by saturation rules\cite{braun2021one}, finding that quality in reflexive thematic analysis derives not from sample size per se, but from transparent, reflexive, and well-justified analytic practice\cite{braun2021saturate}. In this study, the final sample of 15 participants was sufficient to capture diverse perspectives across contexts, with thematic adequacy and depth achieved during analysis. The sections were shaped by identifying common challenges raised in interviews, and aligning those groups with principles of Trustworthy AI from the EU High-Level Expert Group (HLEG), where interviewees provided relevant findings. Sections were combined or created for findings which did not fit into these principles directly. This approach ensured the groupings were data-driven while reflecting established principles. Further research was conducted to explore each challenge, drawing on the latest literature to expand on these findings.

\subsection{Participants}\label{subsec1}
Participants came from a variety of industries which is outlined in table \ref{table1}, with Software as a Service (SAAS) Technology being the most common industry. Participants came from a diverse set of job functions, with technical functions such as data analytics and IT infrastructure/Cyber security being the leading functions. Participants came from a range of industry levels. All participants were based in Ireland and the UK and were well-experienced in their respective fields. Due to location, the findings reflect perspectives shaped by proximity to the EU AI Act and by local organisational and cultural norms. Although many participants worked for multinational companies, with headquarters in the United States of America, China and Europe, their geographic location likely influenced both regulatory awareness and institutional framing of AI governance. The company sizes varied from 10-50 employees to 100k+ employees. All participants were working in organisations where AI is embedded into core business processes, supported by dedicated infrastructure, governance mechanisms, and ongoing operational use, rather than early-stage or pilot experimentation. 

Our sampling strategy deliberately included both technical and non-technical roles to capture a breadth of perspectives across organisational functions. While algorithm engineers, machine learning engineers, and UX researchers were not directly included, our research focus was less on model development and more on governance, compliance, and the organisational embedding of AI. Non-technical roles such as marketing operations and customer support were included to provide additional insights into accountability, user-facing implications, and cross-departmental coordination—critical dimensions of Trustworthy AI implementation. For technically complex issues such as technological robustness or data provenance, findings were interpreted primarily through the lens of participants in data analytics, IT infrastructure, and cyber security roles, whose day-to-day responsibilities directly intersect with these areas. This approach aligns with recommendations in implementation research, where purposeful sampling is used to select participants with relevant knowledge and experience for the study aims, often prioritising variation across roles to capture organisational dynamics rather than technical details of system design \cite{palinkas2015purposeful}.

\begin{table*}[htbp]
  \centering
  \caption{Overview of participants}
  \label{table1}
  \begin{tabular}{|c|c|c|c|c|}
    \hline
    \textbf{ID} & \textbf{Industry Category} & \textbf{Company Size} & \textbf{Job Function} & \textbf{Seniority} \\
    \hline
    P:1 & SaaS Technology & 5k-15k & Marketing Operations & Senior \\
    P:2 & Business Development/BI & 50-500 & Business Development/BI & Mid-Level \\
    P:3 & SaaS Technology & 5k-15k & Product Design & Senior \\
    P:4 & Other & 40k-100k & IT Infrastructure/Cyber Security & Mid-Level \\
    P:5 & SaaS Technology & 1k-5k & Customer Support & Executive \\
    P:6 & SaaS Technology & 500-1k & Data Analytics & Senior \\
    P:7 & SaaS Technology & 1k-5k & IT Infrastructure/Cyber Security & Mid-Level \\
    P:8 & Business Development/BI & 1k-5k & Business Development/BI & Mid-Level \\
    P:9 & Social Media & 40k-100k & IT Infrastructure/Cyber Security & Senior \\
    P:10 & Other & 100k+ & IT Infrastructure/Cyber Security & Mid-Level \\
    P:11 & SaaS Technology & 10-50 & IT Infrastructure/Cyber Security & Executive \\
    P:12 & SaaS Technology & 15k-40k & Data Analytics & Senior \\
    P:13 & Social Media & 40k-100k & Trust \& Safety & Mid-Level \\
    P:14 & Other & 40k-100k & IT Infrastructure/Cyber Security & Mid-Level \\
    P:15 & Other & 40k-100k & Data Analytics & Senior \\
    \hline
  \end{tabular}
\end{table*}

\section{Detailed Research Findings}\label{sec4}
This section organises the findings from our interviews into five sections based on the common themes identified by participants. We extracted the insights provided by participants, highlighting the main points they made, and grouped these insights where they were either consistent or contradictory. These insights were then categorised using the trustworthy AI principles from the HLEG as an initial framework to create high-level thematic groupings. For example, findings related to fairness and bias were generally associated with compliance, so they were placed within a broader category of Accountability, Governance, and Regulatory Compliance. The section titles have been adjusted slightly from the HLEG principles to better align with the specific feedback in each area. Based on the interview results, a comprehensive literature review was conducted for each key finding within the five sections, providing academic context to support the insights gathered. Table \ref{table2} outlines these five sections along with their key findings.

Each subsection includes a table summarising the main findings from the interviews, followed by a discussion of the concerns raised, incorporating both interview insights and academic perspectives. Separate summaries highlight the key takeaways from both sources. Given the focus of this paper on industry perspectives, the findings from industry interviews guided the classification and direction of the subsequent literature review.

To ensure a structured and thorough analysis, each section in this chapter follows a consistent approach: first presenting interview findings, then reviewing relevant literature, and finally synthesizing both perspectives in a concluding discussion.

  \begin{table*}[ht]
  \centering
  \caption{Detailed Findings – Overview of Categories}
  \label{table2}
\begin{tabular}{|p{6cm}|p{8cm}|}
      \hline
      \textbf{Category} & \textbf{Primary Concerns} \\
      \hline
      Accountability, Governance, and Regulatory Compliance & 
      \begin{itemize}
        \item Organisational Structure \& Accountability
        \item Fairness \& Transparency
        \item Challenges with Standards and Regulations
      \end{itemize} \\
      \hline
      Data Management and Quality in AI Systems & 
      \begin{itemize}
        \item Data Quality
        \item Data Provenance, Documentation \& Assessment
      \end{itemize} \\
      \hline
      Human Factors in AI Development and Oversight & 
      \begin{itemize}
        \item Human Oversight and Accountability in AI Systems
        \item Human Bias \& Ethical Implications
        \item People Management \& Organisation Structure Challenges
      \end{itemize} \\
      \hline
      Technological Robustness and Safety & 
      \begin{itemize}
        \item Performance, Reliability and Transparency
        \item Security, Risk \& Trust Concerns
      \end{itemize} \\
      \hline
      Environmental and Societal Impact & 
      \begin{itemize}
        \item Environmental Impact
        \item Societal and Cultural Impact
      \end{itemize} \\
      \hline
    \end{tabular} 
\end{table*}

\subsection{Accountability, Governance, and Regulatory Compliance}\label{subsection1}
This section outlines key challenges and developments in achieving accountability, governance, and regulatory compliance for AI systems. It is divided into three subsections: organisational structure and accountability, which explores internal responsibility gaps; fairness and transparency, which considers biases and the conflict between ethics and business priorities; and challenges with standards and regulations, which discusses the difficulties organisations face in keeping pace with evolving legal and assessment frameworks.

\onecolumn

\begin{longtable}{|p{5cm}|p{10cm}|}
  \caption{Detailed Findings - Accountability, Governance, and Regulatory Compliance} \\
  \hline
  \textbf{Category} & \textbf{Details} \\
  \hline
  \endfirsthead

  \multicolumn{2}{c}{\textit{(Continued from previous page)}} \\
  \hline
  \textbf{Category} & \textbf{Details} \\
  \hline
  \endhead

  \hline
  \multicolumn{2}{|r|}{\textit{(Continued on next page)}} \\
  \endfoot

  \hline
  \endlastfoot

  Organisational Structure \& Accountability & 
  \textbf{Challenges:}
  \begin{itemize}
    \item Siloed Departments and Communication Gaps [P1, P2, P3, P4, P5, P6, P7, P11, P13, P15]
    \item Need for clearer Lines of responsibility [P1, P2, P3, P5, P6, P7, P8, P11, P12, P13]
    \item Having to have blind faith in third-party compliance [P1, P2, P3, P4, P5, P7, P8, P9, P13, P14]
    \item Lack of accountability of companies [P1, P2, P3, P4, P5, P6, P7, P9, P11, P12]
  \end{itemize}
  \textbf{Observations:}
  \begin{itemize}
    \item Formation of AI Councils and Steering Committees [P6, P8, P13]
    \item AI provider/product developer is seen as accountable for the system [P4, P13, P14]
  \end{itemize} \\
  \hline
  Fairness \& Transparency & 
  \textbf{Challenges:}
  \begin{itemize}
    \item Risk of Embedded Biases [P1, P2, P3, P5, P6, P7, P9, P10, P11, P12, P13, P14]
    \item Ethical Implications of AI Decisions are unclear [P1, P2, P3, P7, P8, P11, P13]
    \item Lack of Systematic Bias Testing [P1, P2, P6, P7, P11, P12]
    \item Resource or Time Constraints for Testing and Mitigating Bias [P1, P2, P6, P8, P11, P13]
    \item Explainability of AI Models is a key challenge [P1, P3, P4, P6, P7, P8, P9, P10, P11, P12, P13, P14, P15]
    \item Conflict Between Fairness and Profitability [P2, P5, P6, P7, P9, P13]
  \end{itemize}
  \textbf{Observations:}
  \begin{itemize}
    \item High Reward, Low Risk Business Environment [P2, P6, P7, P9, P13]
    \item Awareness of Biases and Cultural Impacts [P1, P2, P3, P5, P6, P7, P8, P9, P10, P11, P12, P14]
    \item Transparency can increase trust in decisions [P6, P10]
    \item Fairness is not binary. It is measurable on a scale [P2, P3, P11, P13, P14, P15]
    \item Inclusive AI design can lead to wider product innovation [P1]
  \end{itemize} \\
  \hline
  Challenges with Standards and Regulations & 
  \textbf{Challenges:}
  \begin{itemize}
    \item Confusion around the specific impact of the AI Act. [P1, P5, P11, P12, P13]
    \item Lack of AI-Specific Standards for Trustworthy AI [P1, P4, P6, P7, P8, P9, P11, P12, P13, P14, P15]
    \item Rush to Implement AI Without Proper Risk Consideration or Mitigation Processes [P1, P2, P3, P4, P5, P6, P7, P8, P9, P11, P12, P13, P14]
    \item Lack of AI Assessment Tools [P1, P2, P4, P6, P7, P9, P11, P15]
    \item Financial pressure for auditors to sign off compliance [P4, P9]
    \item Over reliance on good faith and assumptions [P1, P2, P3, P4, P5, P6, P7, P8, P9, P13, P14]
    \item Concerns fines are not effective enough or are too low [P2, P3, P5, P7, P9]
    \item ISO standards are not sufficient anymore, real-time, technology-based assessments of AI systems are required [P9, P11, P12]
    \item Need for certifications to be quantifiably validated [P7, P14]
  \end{itemize}
  \textbf{Observations:}
  \begin{itemize}
    \item Belief that regulation is chasing to catch up with technology [P2, P3, P7, P8, P9, P13]
    \item Belief that fines and stopping activities is at least somewhat helpful to police companies [P1, P2, P3, P5, P7, P9, P13]
  \end{itemize} \\
  \hline
\end{longtable}

\twocolumn

\subsubsection{Organisational Structure \& Accountability}\label{subsubsection1}
This section explores how organisational structure influences accountability in AI governance, including the distribution of responsibilities and the effectiveness of internal processes. Interview participants highlighted challenges related to siloed departments, unclear compliance ownership, and the need for interdepartmental collaboration to ensure AI accountability. Below is a detailed summary of the interview findings, subsequent literature review, and a discussion on both.
\paragraph{Interview findings:} Many organisations operate in silos, hindering cross-functional collaboration and creating gaps in compliance implementation. Compliance teams were often disconnected from operational functions, limiting their ability to enforce effective oversight. To combat this, some organisations have started to form interdepartmental AI steering committees with representatives from multiple departments. These committees have a goal to lead the AI strategy and compliance within the organisations. In some cases, professionals noted that when it came to AI compliance, departments were not interested in being responsible for this within their department, and felt another department should make sure they are compliant, in particular as it related to maintaining compliance documentation. Compliance was seen as an administrative task that distracted team members from the primary goals of the department.

There were mixed views on who was responsible for the AI systems Trustworthiness. Some participants felt certain that accountability ultimately sat with the AI provider, regardless of where they sourced their data from, or who built various aspects of the AI system. However, some professionals felt that due to the fragmented nature of AI development, there was a blurred line in terms of the overall responsibility of an AI system. There were additional concerns around working with third-party providers, such as who was responsible if data was purchased to train a model which subsequently created issues for the AI system, or how information could be safely shared as part of an inter-organisational development process. Participants felt they were at the mercy of trusting the compliance of third parties, as even going for a site visit, there was no real way to trust their governance completely. Technical participants (particularly in cyber security and data analytics) emphasised system-level risks and vulnerabilities, whereas non-technical participants, such as those in marketing operations or business development, viewed accountability as the responsibility of external providers or compliance departments. These role-based differences highlight the importance of organisational dialogue across professional functions.

\paragraph{Literature findings:} Researchers argue that mechanisms need to be put in place to allow us to hold decision makers accountable for constitutional and financial consequences including monetary damages or sanctions for companies, as well as broader ethical responsibility (\cite{jensen2016public}).  Legal accountability in AI systems involves assigning responsibility for decisions, managing liability, ensuring transparency, and addressing risks, with evolving frameworks needed to clarify obligations (\cite{uzougbo2024legal}). AI systems are complex and research highlighted the potential for autonomous decision-making create accountability gaps that traditional legal frameworks struggle to address, reinforcing the need for structured regulatory mechanism (\cite{novelli2022ai}). The findings from the interviews were  reiterated in our literature review, which also provided additional insights into the impact of the organisational silos noted by interview participants. These silos are defined as clusters of employees lacking cross-departmental communication, are not necessarily formed due to structural isolation, but instead, interaction patterns are primarily driven by the nature of employees' tasks (\cite{vantaggiato2021breaking}). Siloed departments can inhibit the transfer of information, and contribute to problems in organisations by negatively affecting things such as transparency, accountability, and risk management (\cite{sheaff2017constructing}). These barriers to communication and cooperation can negatively affect organisational efficiency, morale, and innovation. To address these challenges, organisations can adopt a systems approach that emphasises strong leadership, fosters collaboration, and implements structured processes to enhance information flow and teamwork (\cite{bento2020organizational}; \cite{drake2022legal}).

\paragraph{Discussion:} Based on the findings of the interviews and literature review, we can conclude that effective organisational structure needs to enable communication between those with in-depth knowledge of functions involved in AI systems, and those responsible for compliance. structure should be designed to ensure that both parties are well-organised and motivated to collaborate on AI compliance initiatives, fostering alignment and mutual understanding in their efforts to meet regulatory and ethical standards. Stakeholders within departments involved in AI systems such as IT and data departments, should input into the design and maintenance of AI compliance, as those stakeholders have the expertise and knowledge of the systems required to design effective evaluation processes required by the compliance department. Under the EU AI Act, primary responsibility and accountability for an AI system rests with the provider who builds or modifies the system and places it on the market. However, while the legislation specifies this allocation of responsibility, participants in our study reported that accountability processes have not yet been fully adopted or understood within industry, largely due to ongoing uncertainty about the Act’s requirements.

\subsubsection{Fairness and Transparency}\label{subsubsection2}
This section examines the role of fairness and transparency in AI, considering how they are defined, implemented, and the challenges associated with ensuring clarity and equity in AI decision-making. Interviews revealed concerns about bias in AI models, the lack of systematic fairness testing, and the difficulty in balancing business priorities with ethical considerations. Below is a detailed summary of the interview findings, subsequent literature review, and a discussion summing up both.
\paragraph{Interview findings:} AI fairness and transparency emerged as key concerns amongst industry professionals who expressed an increased awareness of biases in models due to the data used for training models. Professionals who worked closely with data and machine learning (ML) models were very aware of how a model can figure out sensitive data, even if it is not explicitly told. There were concerns over risks to embedded bias, with participants noting that biased data was coming out of models and being used to train other models. Additionally, there were concerns over a notable lack of systematic testing for bias. Resource and time constraints contribute to a systematic lack of bias testing and mitigation in organisations. Bias testing was typically informal and initiated by individual interest rather than mandated business practice. Some participants noted that concerns around bias either were not taken or would not be taken seriously due to the conflict with business objectives, primarily profitability. It was also pointed out that fairness concerns a human might pick up on, may not be picked up by a model that’s being asked to optimise for profit. However, one senior data analyst pointed out that even when it was spotted, often it was not prioritised. They explained that there are business targets for profit in sectors like finance and insurance and that this influences the bias of the model. The participant explained that profit needs to be traded off with bias, and it’s usually around feature selection; going on to explain that options are presented to senior leadership, and they make a selection based on their financial targets and projections, with profitability as the primary motivation. Participants noted that efforts to reduce bias for underrepresented groups can introduce trade-offs affecting majority outcomes, while also expressing concern that the growing reliance on synthetic data may amplify bias and undermine fairness and transparency.

A number of participants highlighted the importance of explainability of the AI systems. Two participants noted that introducing transparency in AI systems enabled senior leaders to trust them more and accept AI results quickly. One participant pointed out that while they had internal explainability for the AI models their users directly engaged with, the model explanation was not shared with the users of the platform. Others highlighted a lack of availability of any explainability, even within their organisations. One participant noted that having the transparency between AI provider and client gives the opportunity for the client to input into the model, to ask for additional human in the loop points, and to make changes to the model where needed so that they can trust it. Another participant explained how transparency algorithms such as SHAP (\cite{lundberg2017unified}) and LIME (\cite{ribeiro2016why}) are being used to build trust in AI models internally by enabling them to be explained to management, which increased the trust in the model. Participants also called for transparency around what mechanisms were in place to combat bias, where humans were involved in the process, and what fairness metrics were being used. 

Most of the participants referred to fairness simply as quantifiable bias, limiting their focus to model behavior or dataset imbalances. In doing this, they overlooked how their own design choices and usage decisions may qualitatively contribute to potential negative impacts or risks. However, one participant noted that AI systems should be designed for inclusivity and diversity. They believed that designing AI systems for inclusivity and diversity had additional benefits to the organisation. They gave positive examples of product innovation, such as automatic doors, which were designed for people with disabilities, and self-opening car boots, which were designed by a female product designer returning from maternity leave. Almost half of the participants referenced that compliance should be viewed as a spectrum as opposed to a binary state. Bias, for example, was not seen as something which could be entirely removed without making ineffective the models which had been trained on biased data. Instead, compliance should be viewed on a use-case basis, with varying metrics and levels of adherence required for each situation.

\paragraph{Literature findings:} Researchers reported similar findings to the interviews about the relationship between bias and accuracy in models, noting that although in many instances a trade-off must be made between the two, several researchers proposed methods resulting in what they considered an acceptable balance of bias versus accuracy for their use-cases. Bias can arise from human decisions in the design, implementation, and management of AI systems, with its mitigation requiring a continuous and iterative feedback loop (\cite{suresh2021framework}). Lee (\cite{Lee_2019}) considered bias in the use case of credit lending, proposing considering fairness as variable level of trade-off between competing objectives such as accuracy. Singh et al. (\cite{Singhetal_2021}) developed the Alternate World Index (AWI), a universal fairness metric which they proposed a level of trade-off between fairness and accuracy for credit lending.  Lee and Floridi (\cite{LeeandFloridi_2021}) expanded on the concept of treading fairness not as a binary condition, but instead as a relational trade-off. This transparency allows lenders to justify algorithm choices by balancing financial inclusion and impact on minority groups, while providing regulators and policymakers` with insights to recommend acceptable risk levels. For the use case of recidivism \cite{farayola2023ethics}, Farayola et al. (\cite{farayola2024enhancing}). demonstrated a multi-objective optimisation approach to minimizing bias by examining multiple  fairness-enhancing techniques across different stages of the ML model and examine their impact on the balance between fairness and accuracy. By introducing techniques such as disparate impact remover, adversarial learning, and equalized odds optimisation, the researchers were able to significantly reduce bias with only a minimal cost to the model's accuracy. McCormack and Bendechache (\cite{mccormack2024comprehensive}) classify evaluation criteria for the seven principles of Trustworthy AI originally published by the EU high level expert group, which were also included as non-legally binding guiding principles in the EU AI Act. Their fairness metrics include Group, Individual, Counterfactual, Intersectional, Complex Fairness, and Inclusive Design. They also identify a research gap in AI transparency and propose evaluating systems for model, data, and outcome transparency, emphasizing the importance of visibility into data use and decision-making. The paper stresses the need for standards and processes to address transparency, fairness, and other unknowns in the area of Trustworthy AI. The uncertainty around the future of AI is also echoed by Floridi (\cite{floridi2020ai}) who refers to speculation on the hype of AI as the wild west of “what if” scenarios. They say this is impacted by media oscillating from an AI utopia to an AI doomsday scenario and propose that AI be viewed as a normal technology rather than a miracle or plague. The paper calls for more philosophical thought into ethics, and consideration around what technology is being developed and its potential impacts. The paper argues that ethical frameworks and principles to underpin AI technology do exist, but states a need for more thought into how AI will fit in the developing human-technology relationship. 

\paragraph{Discussion:} AI is being developed at such a rapid pace that existing AI systems frequently perpetuate bias and discrimination, which can be introduced at various stages of the AI system(\cite{suresh2021framework}, raising significant concerns about the unknown ethical implications. There is a lack of transparency and established processes for evaluating AI systems' fairness. Moreover, business objectives, particularly profit, are often prioritised over societal and ethical considerations. While technology to enhance transparency and fairness has already been developed, it is not yet sufficiently adopted or audited by regulatory bodies. There is an urgent need for the standardisation and evaluation of fairness and transparency practices in AI systems, particularly concerning the trade-offs between profit and fairness in decision-making processes.

\subsubsection{Challenges}\label{subsubsection3}

This section discusses the complexities of regulating AI, the difficulties in assessing compliance, and the evolving landscape of governance frameworks. Interview participants expressed frustration over unclear regulatory expectations, the reactive nature of compliance measures, and the challenges of holding AI developers accountable. Below is a detailed summary of the interview findings, subsequent literature review, and a discussion on both.

\paragraph{Interview findings:} Professionals found several challenges when it came to regulations and assessment for Trustworthy AI. They noted that companies were currently operating with very little accountability for their AI activities, with several participants believing that regulation was struggling to catch up with technology. Professionals referred to the current AI development environment as the “Wild West” due to the high ratio of reward versus risk for companies. Several professionals felt this was because stealing data or engaging in unethical AI practices did not currently have sufficient repercussions in place. One participant described an attitude of entitlement within organisations when it came to taking data, even illegally for the sake of AI innovation. Although many professionals felt that fines were somewhat effective in holding businesses accountable, ultimately, they noted that fines are often seen as an acceptable risk by companies. Prohibiting companies from engaging in activities due to breaches was suggested by some professionals as a better way to hold organisations accountable, as it would have more impact on profitability. Participants felt that companies would tend to opt for the bare minimum for legal compliance.
When it came to implementation, professionals faced challenges in putting Trustworthy AI practices in place. One issue was a lack of tools available to easily assess AI systems against a  trustworthy AI standard. While professionals felt that the AI Act would be significant to organisations, they noted that there was confusion around its implications, along with a lack of resources and clarity such as those currently available for GDPR compliance. Some professionals identified bias but felt unable to act due to anticipated resistance or previous dismissal by leadership.

The industry’s reliance on good faith, both in self-regulation and third-party verification, was seen as a key vulnerability in compliance enforcement. Approaches to compliance were typically responsive “band-aid” solutions designed to meet minimum requirements stipulated by the organisations compliance department. Employees are both pressured and financially incentivised to make sure they assess systems in a way that would pass their compliance processes. They noted that company bonuses were often dependent on achieving company objectives such as compliance certification for their systems. Due to the over-reliance on good faith, doing the bare minimum or even lying to get certification, to achieve this financial reward was both possible and incentivised. This was the case for both internal employees and external auditors, who in some cases would have a prior discussions around compliance for certifications such as ISO27001\cite{iso27001}, before the official findings of the report were published. This allowed the company the opportunity to liaise with the auditor before any official report maintained for an audit trail was completed. Paying auditors for certification created perceived conflicts of interest, raising doubts about the objectivity and integrity of compliance processes. The existing processes were largely around maintaining documentation, and less about the quality of the documentation. The ability to make human judgement calls around what documentation was considered sufficient added ambiguity into the audit process. This problem was echoed by participants who noted a need for certifications which were able to be quantifiably validated. Participants noted that there has to be technology and tools that go hand in hand with design, development, and deployment that collect data in real-time if possible.

\paragraph{Literature findings:} The literature reports similar concerns around several issues relating to regulations and assessment for trustworthy AI. Jobin et al. (\cite{jobin2019global}) highlighted concerns about lack of accountability in AI development. They found that companies were able to bypass ethical requirements due to insufficient regulatory enforcement or the use of high-level soft-policy approaches. They describe uncertainty around how ethical principles should be evaluated, as there was no clear process for enforcing oversight. In particular, they note a gap in the ability of regulators to prioritise conflicting ethical principles. They proposed twenty-two approaches to effecting change in TAI in companies under the four high-level classifications of social engagement, soft policy, economic incentives and regulation and audits. A recent paper by Diaz Rodríguez et al. (\cite{diaz2023connecting}) reports that accountability issues are still prominent, calling for enhanced regulation and oversight mechanisms. The researchers highlighted the importance of developing social and ethical standards that could first be implemented in the design and construction of systems and subsequently used to assess those systems for their conformity. Percy et al. (\cite{percy2021accountability}) highlighted the importance of both external accreditation and internal audits to foster trust and transparency and establish a balanced ecosystem. They argued that reliance solely on high-level ethical principles or external regulations is insufficient for accountability. Using the example of gender bias in gambling, they showed that gender bias can be reduced, albeit with a cost to overall model accuracy. Due to the variability in AI systems, they concluded that industry-specific guidelines were essential for accountability in addition to internal audits and explainability processes. They expressed the concern that without improved governance, companies may continue to bypass ethical responsibilities. The researchers also noted that a variety of tools and techniques have been developed and published to help machine learning developers to implement ethical principles at various stages of the development process, however there is no agreement on how these should be measured or enforced.

Ewuga et al. (\cite{ewuga2023iso}) investigated the implementation and effectiveness of the ISO27001, a risk based framework for information security management systems prominently used as part data protection in organisations. Their research, which looked at the banking sector, showed many benefits, including improved risk management, incident response, and cultivating a security-aware culture. It uses a Plan-Do-Check-Act approach which requires continuous monitoring and improvement and includes ethical considerations such as customer consent and balancing data subjects' privacy with system security and transparency. The researchers found that banks have dynamic cyber threats which can often lack evaluation metrics and indicators. In their discussion on addressing future challenges with emerging technologies including AI and ML, the researchers call for the development of specific and stringent cybersecurity standards for these evolving threats.
Kamil et al. (\cite{Kamil2023information}) found that there was pressure on employees within organisations to maintain compliance with ISO27001 which sometimes led to them becoming bad actors. The pressure felt by employees could result in unintended risks such as bypassing security measures or exploiting system gaps to pass audits. The certification was seen as somewhat devalued as some clients were not accepting this standard alone but required multiple standards to feel assured of their commitment to information security.
Fontrodona et al. (\cite{fontrodona2013relation}) researched the relationship between innovation and ethics, finding that they are closely interconnected. While innovation is about exploring the possibilities, ethics provides a framework to ensure this progress is aligned with high-level societal principles. The authors noted that innovation that is not rooted in ethical considerations can lead to harmful societal or environmental consequences. 

\paragraph{Discussion:} Both the literature and interviews highlight significant challenges related to accountability in AI systems. Current accountability standards, which rely on risk-based frameworks and manual processes such as checklists, are regarded as insufficient. There is a clear need for quantifiable standards and metrics that can be continuously monitored throughout the lifecycle of AI systems. Regulators often place undue reliance on good faith, with organisations depending heavily on their own auditors or certification bodies. Given that these auditors and certification companies are paid for their services, they may be financially incentivized to act in the best interest of the business, which can undermine their ability to hold organisations accountable.

Within organisations, employees may face pressure, leading them to take shortcuts or exploit vulnerabilities to pass compliance audits. While tools and techniques exist to integrate ethical principles at various stages of the AI development process, implementing Trusted AI (TAI) is often hindered by conflicting business goals. For example, in sectors such as banking and insurance, machine learning models can be quantitatively assessed for bias, but efforts to reduce bias usually come with a trade-off in accuracy, which negatively impacts profitability and is thus often deprioritised.

This conflict between business objectives and ethical considerations results in organisations setting acceptable levels of bias in machine learning models, leading to profit-driven models rather than fair ones. The lack of ethical considerations in the innovation and development of AI technology is already contributing to harmful societal consequences. To enable effective ethical accountability, independent third-party accreditation bodies must establish Trustworthy AI evaluation criteria tailored to specific use cases at a sector level, incorporating standardised metrics that must be adhered to.

\subsection{Data Management and Quality }\label{subsection2}
\begin{table*}[htbp]
  \centering
  \caption{Detailed Findings - Data Management and Quality}
\begin{tabular}{|p{5cm}|p{10cm}|}
      \hline
      \textbf{Category} & \textbf{Details} \\
      \hline
      Data Quality & 
      \textbf{Challenges:}
      \begin{itemize}
        \item Issues with accuracy and trustworthiness [P1, P2, P3, P4, P5, P6, P7, P8, P9, P10, P11, P12, P13, P14, P15]
        \item Inconsistent Data Entry [P1, P2, P3, P4, P7, P8, P13, P15]
        \item Overreliance, validity or bias concerns around Synthetic Data [P3, P7, P11]
        \item Concerns with data vendors faking consent [P4, P9, P10, P11]
      \end{itemize} \\
      \hline
      Data Provenance, Documentation \& Assessment &
      \textbf{Challenges:}
      \begin{itemize}
        \item Challenges in Maintaining Up-to-Date Documentation [P1, P2, P3, P4, P6, P7, P8, P9, P12, P13, P15]
      \end{itemize}
      \textbf{Observation:}
      \begin{itemize}
        \item Benefits to companies with good Data Lineage Tracking [P1, P2, P8, P9, P13, P15]
      \end{itemize} \\
      \hline
    \end{tabular} 
\end{table*}

\subsubsection{Data Quality}
This section addresses the importance of data quality in AI systems, including the factors that impact data reliability and the implications of poor data practices. Interviewees frequently cited issues with inconsistent data entry, vendor misrepresentation of data consent, and the risks posed by low-quality or biased data feeding AI models. Below is a detailed summary of the interview findings, subsequent literature review, and a discussion on both.

\paragraph{Interview findings:} Insights in this section were drawn primarily from participants in technical roles relating more closely to data quality (data analytics, IT infrastructure, cyber security). While non-technical roles provided complementary perspectives, the technical participants’ expertise informed the more detailed aspects of these themes. All fifteen participants highlighted issues with data accuracy and reliability. The most mentioned issue was around inaccurate model outputs resulting from low-quality or misunderstood data. Participants emphasised that poor-quality input data leads directly to unreliable model outputs, highlighting a persistent ‘rubbish in, rubbish out’ issue. There were several ethical concerns flagged around biased data leading to biased outcomes that reflected social prejudices. Participants flagged challenges around inconsistent data entry practices, either from human-entered data or fields that were not common across departments and subsequently misinterpreted. One participant who worked as a lead data scientist explained the benefits of assigning quality thresholds to input data so it could automatically be flagged for changes affecting quality. They explained that the model performance metrics which were also tracked, took some time to pick up the mistakes, versus it being spotted at source. Despite data quality being more critical to model success, it was frequently deprioritised in favour of advancing model development. Additionally, when the error is flagged at the input data, it makes it easier to fix as the source of the issue is immediately identified. 

Concerns were raised by four participants around data acquisition and cleaning practices, specifically around data vendors misrepresenting consent. It was pointed out that vendors will claim to be the originator of the data, when they are not. Additionally, it was mentioned that there was no way to verify if sets of data with and without consent were merged and disguised as fully GDPR-compliant data when they were not. The lack of transparency in data sourcing from brokers could lead to problems for purchasers who are relying on good faith that the data is being represented truthfully. There were questions about how compliant an AI builder could be if they bought data to train their model, which later turned out to have been misrepresented by the data vendor. 

Three participants flagged concerns around the increasing use of synthetic data. The primary concern was that the data would have increasing bias and validity issues affecting the model’s performance. This cycle was referred to as a chicken and egg problem.

\paragraph{Literature findings:} The literature also highlighted that data quality is an essential part of the performance of machine learning models, and that it was also often under-prioritised versus model development, along with concerns raised around synthetic data and data provenance. Gupta et al. (\cite{gupta2021data}) found there are many issues around data collection and processing affecting the reliability and accuracy of models. Researchers highlighted the amount of time spent by data scientists on debugging data, citing the importance of well-defined quality metrics as a way to save time. The paper also proposes a series of quality metrics for various aspects and types of AI systems. Ragineni (\cite{rangineni2023analysis}) also highlighted the importance of data quality and noted the issues that can occur during the machine learning pipeline. The paper outlines a comprehensive data cleaning process, and also proposes additional data quality dimensions to address challenges around software, bias and legal and ethical aspects. Priestley et al. (\cite{priestley2023survey}) argued that data quality means different things to different people. To ensure that data quality was contextual, they proposed mapping traditional criteria such as accuracy and consistency to specific stages of the ML lifecycle. The paper offers additional practical insights and advice around stakeholder needs, data quality across complex ecosystems and data quality management. 

Sambasivan et al. (\cite{sambasivan2021everyone}) noted that model development was often overvalued versus data quality, and this led to negative outcomes. They found a need for cross-domain collaboration, which was a challenging task due to poorly maintained documentation, variability in incentives and insufficient domain expertise, which contributed to project failures. The researchers also noted that data collectors and vendors involved in AI projects can be under-resourced or lack proper training. This can lead to conflicting agendas where field workers may either fabricate data or fail to collect adequate data due to a lack of motivation or understanding of the importance. Additionally, Morey et al. (\cite{morey2015customer}) found a lack of transparency in the collection of personal data which caused concerns with customers. They found that some organisations were exploiting data for short-term gain but that was not a sustainable approach due to growing customer demand for control of their data. Hug (\cite{huq2021public}) discussed the manipulation of data from individuals for profit, noting that companies circulate data without transparent consent from individuals. Whitney and Norman (\cite{whitney2024real}) explore the area of circumventing consent, where companies take data without consent and manipulate it to sidestep actually obtaining consent. The authors detailed the practice of organisations taking personal data and using it to generate synthetic data, obscuring the data’s origin. 

Jordon et al. (\cite{jordon2022synthetic}) also found that synthetic data comes with a number of risks, including privacy and bias concerns. While datasets can be augmented to create de-biased synthetic datasets, fixing the inherent bias in the original data, the process comes with its own set of risks to create new problems, which need to be closely monitored. Joshi et al. (\cite{joshi2024synthetic}) highlight concerns with synthetic data, such as leakage, dataset diversity, and representation fidelity, along with approaches to mitigate these risks. 

Delacroix et al. (\cite{delacroix2019bottom}) highlighted the power imbalance between data controllers and data subjects, in part due to an inability to effectively implement governance measures such as GDPR. They proposed a bottom-up data trust solution in which data subjects could pool their data under a fiduciary structure to offer collective empowerment over their data.

\paragraph{Discussion:} Both the interviews and literature showed the importance of data quality, in particular, the benefits of having well-defined data quality metrics to ensure consistent model performance. Unethical concerns were raised around current practices of biased and manipulated data leading to negative outcomes. Synthetic data is an area of growing concern for similar reasons including legalities of consent, prejudice and validity concerns. Looking forward, both traditional data quality dimensions and data cleaning processes need to be updated and embraced by industry to support better data governance—ultimately leading to greater transparency and improved performance of AI systems. There are also calls for changes in the structure of data ownership, with a shift towards increasing control for data subjects over how their data is used.

\subsubsection{Data Provenance}
This section considers the significance of data provenance, focusing on how organisations track, document, and verify the origins of data used in AI models. Interviewees highlighted difficulties in maintaining up-to-date documentation, the time wasted tracking down data sources, and the negative impact of poor data lineage on AI reliability. Below is a detailed summary of the interview findings, subsequent literature review, and a discussion on both.

\paragraph{Interview findings:} Six participants mentioned the benefits of maintaining good data lineage within organisations. Participants noted that time is frequently wasted chasing the source or proper meaning of data. This can include just emails or phone calls to colleagues or, on occasion, actually chasing down the people on the floor who collected the data to understand the meaning and context of various fields. Participants noted that models had been hastily built in companies in the AI rush, which subsequently didn’t work due to the model builder not understanding or making false assumptions about the data used to train the model. Additional benefits of good data tracking included ensuring data integrity, compliance, improved transparency and improved trust in the data. The benefits of tracking data through the company, including data flow diagrams and real-time accurate documentation, are also highlighted. Many participants noted that a lot of their answers around data could be found in internal repositories. Despite this, they mentioned chasing additional information about data to assist with their projects. This was often due to documentation either not being complete or being out of date. This was particularly the case in dynamic environments where both the data and the AI models are evolving. There was a fragmentation of responsibilities, with people finding the task of continuously updating documentation to be labour-intensive admin work they did not want to do. It was seen as checklist-type work, which often did not keep pace with product deployments. The fragmented development also created challenges in ensuring consistent data oversight, in particular when there was a lack of integrated systems.

\paragraph{Literature findings:} The findings in the interviews align with the existing literature on data quality and data provenance in AI systems. Werder et al. (\cite{werder2022establishing})highlighted the importance of data provenance for mitigating bias and making AI systems more responsible. They note that organisations see data provenance as a compliance mandate rather than as part of an organisational commitment to developing responsible AI systems. They found that when organisations developed practices such as the automated ones recommended in their paper, they had better long-term outcomes for the organisation, particularly in data-driven development and AI engineering. Laine et al. \cite{laine2024ethics} also notes the importance of data provenance for mitigating bias and promoting accountability and transparency in AI systems, in particular in relation to decision pipelines. They also note that fragmented documentation with missing information leads to challenges for businesses during fairness and compliance audits.
Solomon and Brown \cite{solomon2021influence} found that organisational culture played a significant role in informational security subculture, affecting how employees followed processes such as good data provenance. They found that a top-down approach, including the monitoring of compliance processes and ensuring employees were aware of their accountability in those processes, was important. Additionally, good data provenance processes can offer additional benefits such as also smoothing transitions between employees, in particular in instances such as company downsizing \cite{mclachlan2022developing}.

\paragraph{Discussion:} Data provenance, when implemented as a band-aid style compliance solution, causes friction and creates problems that cascade throughout the organisation. Examples of this include both labour-intensive tasks around chasing data sources and meaning and the building of AI products that don’t work due to data misunderstandings. Businesses that fail to create both good data processes and a culture of data compliance will see increasingly negative organisational outcomes as AI becomes more commonplace. Strong data provenance practices that are rooted in best practices, in particular those that include automation, will offer significant benefits to organisations.

\subsection{Human Agency and Oversight}

\begin{table*}[htbp]   
\centering   
\caption{Detailed Findings – Human Agency and Oversight} 
    \begin{tabular}{|p{5cm}|p{10cm}|}
      \hline
      \textbf{Category} & \textbf{Details} \\
      \hline
      Human Agency \& Oversight & 
      \textbf{Challenges:}
      \begin{itemize}
        \item Not enough humans-in-the-loop in AI systems [P1, P2, P3, P4, P8, P10, P13, P15]
        \item Risks of over automation [P1, P2, P5, P6, P7, P8, P10, P12, P14, P15]
        \item Need for ongoing monitoring [P1, P5, P9, P11, P12, P13]
        \item Humans need to be able to understand model decisions [P1, P3, P4, P6, P7, P8, P9, P10, P11, P12, P13, P14, P15]
        \item There is a need for human checks at certain points [P1, P3, P4, P5, P8, P13, P14, P15]
        \item Concerns that human-in-the-loop could introduce bias [P11, P12]
        \item Human decisions need accountability [P11, P12, P13]
        \item Organisational structural changes needed for oversight [P1, P2, P11]
        \item Call for mandated oversight roles in companies [P1, P2]
        \item Concerns around education of professionals [P1, P2, P8, P10]
      \end{itemize}
      \textbf{Observations:}
      \begin{itemize}
        \item Humans can help mitigate known and unknown ethical implications in system design or data collection, processing \& testing [P1, P4, P5, P6, P7, P10, P15]
      \end{itemize} \\
      \hline
\end{tabular}
\end{table*}
  
This section explores the role of human oversight in AI systems, assessing the balance between automation and human decision-making, as well as the need for accountability. Interviews highlighted concerns about insufficient human oversight, gaps in AI education among professionals, and conflicts of interest in human decision-making. Below is a detailed summary of the interview findings, subsequent literature review, and a discussion on both.

\paragraph{Interview findings:} Participants had concerns there might not be enough humans in the loop in systems. There were concerns around over-automation of systems, in particular without proper monitoring, robustness and safety being built into them. One participant noted that they expect their organisation’s tools not to allow them to do anything that isn’t compliant, and that compliance should be inherently built into the system, but this was not always the case. There were also concerns that even when things are GDPR compliant, we still didn’t know enough about the potential ethical implications. A number of participants felt there is a need for people to be in the AI process of making decisions at mandatory gates. One participant noted that AI could check AI in the future, but at this stage, humans should be monitoring these checks due to AI verification tools not being available. One participant gave the example of pilots who have certain thresholds and points that they are required to take over, noting that this should be the same to ensure the safety of all AI systems. From a usability perspective, one participant noted that people sometimes don’t realise they are dealing with an AI, so you need humans to be available when issues arise.
Participants also noted that the AI can miss things due to context or just not being advanced enough, and this is something humans are able to pick up on. One participant noted that humans in the loop should not be incentivised to choose profit, stating that if a human is a shareholder, they will not be able to make an unbiased decision between profit and ethics. Two participants pointed out that humans in the loop can introduce their own biases into the system also.
There were concerns about the educational training of humans involved in AI systems in areas such as poor data literacy or professionals having fragmented knowledge of AI systems. One participant commented that issues are arising because a lot of models are being built by engineers who don’t understand the full process, how data is collected and processed and how it’s going to be used by end users. A second participant noted a similar focus on the model build, stating that data scientists are directed to focus on math and algorithms, leading to a significant gap in education around the data aspects. Another participant highlighted the need for the addition of specific people or tools to ensure compliance with AI systems. A fourth participant commented on the lack of education of employees due to the rush to implement AI in their organisation. They commented that this was happening without educating employees on the risks, including the unknown risks involved, so that they could help mitigate these.
A number of participants noted a need for change in organisational structure, with this already starting to happen in many organisations. AI teams or AI councils are being retrospectively set up to look after privacy and governance. Some participants called for new roles or AI compliance departments, such as AI equivalent to the GDPR-mandated data protection officer, questioning why this wasn’t included in the act. One participant noted that smaller companies who now have access to this powerful technology may not have the resources to ensure compliance with it. There was also a reported disconnect between data controllers and end users. One participant said that instead of replacing departments or roles in companies with AI, the roles should be halved and become AI-assisted with enough human checks in place to mitigate risks.
The explainability of models to humans, as discussed in section \ref{subsubsection2} in this paper was also flagged as important by participants. One participant highlighted that proper human oversight is only possible when the humans understand what is being done, the input, output and the steps in between.
Participants noted that human decisions need accountability. Companies that are incentivised to favour profit over bias need to be made accountable for their decisions. Additionally, humans who could introduce bias into systems interviews, needs to be transparent so that it can be monitored for quality checks and auditing also. It was noted that monitoring of AI systems would have require ongoing accountability and monitoring, an AI tool that was originally safe to use could be compromised by a bad actor gaming it or introducing bias. One participant described their organisational GDPR implementation as a big project initially, but once it was in place, ongoing monitoring and training were sufficient.

\paragraph{Literature findings:} The literature also found that human oversight in AI systems must be carefully designed, with particular focus on transparency, ongoing monitoring, and the level of training and motivation of those involved. Enqvist \cite{enqvist2023human}, noted that human oversight can be designed in many ways, with many outcome variabilities based on the selection of which processes, which person and which time the process is overseen. They also pointed out that the oversight should be designed with attention given to transparency and the mandates and working conditions of the human oversight to more effectively counterbalance the risks they are trying to mitigate. They highlighted that human monitoring should be ongoing to address risks and biases instead of reactive. The paper also discusses the human-centric approach discussed in laws such as the AI Act, noting that this mandated oversight requirement is being written into law with no precedent for what the human-centric approach applies, and a lot of room left for interpretation by AI providers. 

Automation bias leads humans to uncritically rely on automated systems, even when faced with contradictory information, showing that AI can introduce new biases through the way its outputs are interpreted rather than simply eliminating existing ones \cite{alon2023human}.
Human oversight is a potential safeguard to mitigate some risks in AI systems in the hope that humans may be better at including ethical considerations and adhering to social norms in decision-making processes \cite{sterz2024quest}. However, this paper also noted that humans may come to rely too heavily on the AI outputs and be influenced by them, or else counter them unfairly and introduce their own biases. They noted that any human in the loop would have to be trained with sufficient knowledge about the risk and how to mitigate it, also calling attention to the accountability of having an appropriate person in the loop, stating that effectiveness requires both moral responsibility and fitting intentions. The researchers propose an approach to effective human oversight under three categories: the technical design of the system, individual characteristics of oversight persons, and the environmental circumstances in which oversight occurs.

Researchers discuss what an acceptable standard of care, a norm-based governance with legal implications, would entail for AI systems \cite{cihon2024chilling}. They found AI providers need to take reasonable actions and precautions to ensure no resulting harm, tort or regulatory liability, and this includes implementing human oversight. The paper states that human oversight includes both the information to responsibly use an AI agent and control it during operation as opposed to autonomy, which is at the other end of the spectrum. They state that for a use case, such as when a driver should take over driving a vehicle, the appropriate level of oversight versus autonomy must be selected, which reflects the advances in capabilities and safety. They argue that AI systems which pose challenges to effective human oversight could be seen as defective products under the law. The paper states that more research into the human-AI relationship will help determine an appropriate standard of care for specific use cases.

\paragraph{Discussion:} There is a clear need for sufficient human oversight in order for AI systems to be considered safe and ethical. The legal implications of what an appropriate level of oversight or standard of care might look like is currently unclear. The literature and regulations in this area state that humans providing oversight or operating AI systems should have sufficient training. However, industry participants noted concerns that sufficient training was not in place currently. There is a shared concern in the literature and industry that over-automation and lack of human control could occur either through lack of human oversight or through human decision-making either being influenced or wrongly influencing AI systems. This could result from a lack of transparency, lack of education, or human oversight by individuals with conflicting agendas regarding the ethical integrity of the system. Despite these concerns, human oversight is seen as integral to ensuring fairness and safety in AI systems. There are concerns that achieving legal compliance with AI standards will not result in meeting ethical standards. Accountability and transparency of human involvement in AI systems involved in high-risk decision-making is seen as essential. Human agency versus human oversight is a balance that must be struck for each use case. It is likely that significant organisational changes in structure will result from the implementation of AI technology with sufficient human involvement. 

\subsection{Technological Robustness and Safety}
This section focuses on the technical resilience and security of AI systems, considering issues with performance, reliability, transparency, evolving cyber threats, and the need for new standards and risk mitigation measures to maintain trust and safety.

\begin{table*}[htbp]   
\centering   
\caption{Detailed Findings – Technological Robustness and Safety} 
    \begin{tabular}{|p{5cm}|p{10cm}|}
      \hline
      \textbf{Category} & \textbf{Details} \\
      \hline
      Performance, Reliability and Transparency &
      \textbf{Challenges:}
      \begin{itemize}
        \item Issues with model Reliability and Accuracy [P1, P3, P8, P9, P10, P11, P12, P13]
        \item Issues with hallucinations [P3, P14]
        \item Risk of overreliance on AI [P2, P3]
      \end{itemize}
      \textbf{Observations:}
      \begin{itemize}
        \item Automated data and model transparency initiatives improved organisational trust in AI [P6, P15]
        \item Importance of reproducibility [P9, P13, P14]
      \end{itemize} \\
      \hline
      Security, Risk \& Trust Concerns &
      \textbf{Challenges:}
      \begin{itemize}
        \item Evolving security needs [P1, P7, P11]
        \item Sensitive data leaking into model [P3, P9, P10]
        \item Data Poisoning and Adversarial Attacks [P9, P11, P14]
        \item Increase in volume of attacks [P1, P2, P9, P11]
        \item Need for strong risk mitigation [P1, P3, P4, P9, P10]
        \item Concerns current assessments like ISO27001 were not sufficient for AI [P9, P10]
        \item Found existing security measures sufficient for at least some aspects of AI systems [P3, P4, P5]
      \end{itemize}
      \textbf{Observations:}
      \begin{itemize}
        \item Fear of the unknowns [P5, P7, P11]
      \end{itemize} \\
      \hline
    \end{tabular} 
\end{table*}
  
\subsubsection{Performance, Reliability and Transparency}
This section examines the technical robustness of AI systems, including the factors that influence performance, reliability, and the importance of transparency in AI operations. Interviewees raised concerns about model accuracy, the impact of data quality on performance, and the need for transparency to build trust in AI decisions. Below is a detailed summary of the interview findings, subsequent literature review, and a discussion on both.

\paragraph{Interview findings:}
Insights in this section were drawn primarily from participants in technical roles (data analytics, IT infrastructure, cyber security), whose responsibilities involved direct engagement with system robustness, with non-technical roles providing complementary perspectives. Data accuracy is essential in robust AI systems. Participants commented on the risks of low-quality or poorly structured data, giving examples of how they had gotten inaccurate outputs or poor model performance as a result. Participants highlighted challenges in people not understanding the data, not being data literate, and also having issues chasing back data sources to find out what data really means and where it was acquired so they can use it effectively in models. Participants felt that organisations needed to focus more on model accuracy and robustness. Many participants highlighted the impact of conflicting goals, such as improving bias and having negative outcomes on model performance.

Two participants highlighted the negative impact of hallucinations, which they said was also a result of poor data or, at times, out-of-date data. One participant commented that the quality of the ML model can be compromised when models are fed chunks of data from various sources that conflict with or confuse the model, causing hallucinations. Another participant noted that hallucinations could frequently occur in specific topics that haven’t been well-trained on the model, leading to confident hallucination responses to fill the gaps.

The positive benefits of introducing automated techniques to improve transparency was highlighted by participants. One participant described how introducing rule-based dashboards for data quality fields such as completeness, validity, reasonableness, and consistency improved both their model performance and gave organisation leaders confidence. Another participant described how the introduction of model transparency techniques gave confidence to senior leaders in AI decisions. 
Reproducibility was highlighted as important, with participants noting the importance of being able to do things such as going back and checking data when there are variances in figures and being able to revert to the previous state if something goes wrong in processes that are in place. One participant noted that AI assessments should be repeatable by software in heightened security risk situations.
Overreliance on AI systems was also listed as a risk by some participants. One participant explained how employees already expect systems to be fully compliant and not allow them to do anything they shouldn’t.
\paragraph{Literature findings:} The literature shows a number of common themes with research participants’ comments in the area of technical robustness and safety. A technical report by the Joint Research Centre and the European Commission’s Science and Knowledge Service drew attention to the importance of reproducibility, risks of overreliance on AI, and discussed issues with model reliability and accuracy\cite{hamon2020robustness}. The report highlights two key aspects of evaluating an AI system’s reliability: performance and vulnerabilities. They explain that poor performance would be a model that cannot perform well in a task that is considered normal for humans, and vulnerabilities would be performance that malfunctions in specific conditions, either unintentionally or through intentional malicious provocation. They discuss how model performance can vary when other features are added, such as introducing interpretability, which requires a trade-off with accuracy. They highlight the importance of reproducibility for AI reliability. The report highlights the importance of external data validation to help avoid overfitting and improve model performance in diverse scenarios. The report highlights the risk of blind trust in AI systems, which can lead to overreliance without sufficient knowledge of the system, which could have negative outcomes like errors or abuse of AI systems. 
The literature also reports a correlation between transparency and user trust. Transparency is required for building trust in Human-Centred Artificial Intelligence (HCAI), stating that the lack of procedures for investigating incidents, along with the lack of specifications of standards for black-box technology is inhibiting trust in those systems (\cite{falco2021governing}). This paper also proposes automated real-time risk evaluations to provide transparency for stakeholders. Additionally, increasing either transparency or control over decision-making in AI systems correlates positively with user trust in the system (\cite{mccormack2024comprehensive}). 

\paragraph{Discussion:} The insights from the study participants, and the supporting literature emphasise critical role of data accuracy and data management processes in ensuring technical robustness and safety in AI systems. Poor data literacy and lack of transparency or lack of knowledge about AI systems create risks for model performance and hallucinations. The possibility of negative outcomes from overreliance on AI systems was also expressed. Additionally, ethical challenges between who decides the appropriate trade-off between bias and accuracy were highlighted, with an expressed need for standard practices to be developed to increase transparency. Increasing transparency was shown to increase user trust and AI adoption within organisations, giving confidence to leadership in AI models. Alongside transparency, reproducibility was flagged as essential in ensuring the reliability of AI systems.

\subsubsection{Security, Risk and Trust Concerns}
This section considers the security challenges associated with AI, including emerging risks, vulnerabilities, and the measures required to build trust in AI systems. Interviews revealed growing concerns over adversarial attacks, data leakage, and the inadequacy of existing security frameworks in addressing AI-specific threats. Below is a detailed summary of the interview findings, subsequent literature review, and a discussion on both.

\paragraph{Interview findings:} Participants noted there were evolving security measures with AI, which some felt wasn’t thought out enough, and that security wasn’t moving as quickly as the technology, with new classes of security threats emerging all the time. One participant explained that when they create a model and apply it to an environment, it mathematically makes sense, but it must be validated, which is difficult when data sets fluctuate and need to be aggregated. Another participant said that AI tools are so much more powerful than people realise, and they are being developed without technical robustness and safety being built in.
Participants noted risks around malicious attacks. One participant noted that processes for data acquisition, cleaning, and transformation are very light in determining whether that data has malicious intent. Another noted the importance of securing not just the model itself but the risks of injecting malware into data and data poisoning. Another noted that AI systems need to be protected from development pipelines, which can be interfered with, similar to supply chain attacks. 

Sensitive data leaking through a model was a concern for some participants, particularly one participant who was in the process of incorporating AI into their business intelligence product, which would have sensitive competitor data on it; it was clear that it could not be leaked between clients. Another participant explained that AI can scrub data but that people can convince it to leak that data.
Four participants noted that there was an increase in the volume of cyber-attacks. Some noted that this was because as AI lowered the bar to entry for hackers. One participant commented that hackers are no longer required to know an obscure programming language to launch good cyber-attacks; and another participant commented how individuals and companies are using AI to launch an increased volume of attacks. Another reason given by three participants was the increased surface area for attacks, with one participant noting that every time they transmit data, it’s a risk for security and another commenting that they open up their models for other APIs and create multiple entry points, making it harder to protect. Data accuracy was cited as important in cyber security by one participant who noted that AI and Large Language models (LLMs) have created a huge new territory in cyber security with new risks.  

Risk mitigation was a key concern with participants highlighting the need for proper procedures and techniques in place. One participant said that risk level needs to be evaluated and made relevant to the level of oversight. One participant highlighted the importance of having data breach guidelines and risk mitigation processes in place with the provider to bring back the trust and repair any situation where something went wrong. Internal mistakes from AI builders to other teams were also highlighted as an important area to have proper checks in place. Another participant explained that the key thing for privacy is data minimisation, but with data being so valuable, companies are collecting it without techniques for data privacy being well implemented. They explained that sometimes data privacy is implemented in sections and can later be combined with other data that was purchased. So, in itself, it’s anonymous, but when everyone has a piece of that jigsaw, they can put that together. One participant noted the benefits of looking after their own data, explaining they had implemented a very strict system for moving data around the company, which could only be accessed through VPN, and even then is highly regulated by legal and only on request.
Participants highlighted a number of issues with AI assessments. One participant explained that the format of the current ISO standards is not sufficient for evaluating AI systems as they don’t assess anything in a mathematically proven way. They also explained that the current ISO27001 standard should not be used for AI as it is a point-in-time standard and framework, and AI is fundamentally not assessable for that control. They said that the standard doesn’t address AI-specific threats or threat vectors, that it has not been updated for AI, and cannot be updated in its current form. Another participant explained that the lack of verifiability of systems created trust issues when selecting providers to look after data. Stating that even though they can test their own security measures, the uncertainty generated by back to lack of actual assessments of data centres means that even if they physically go and look at data centres and examine systems, they still have to just trust that it will be secure. Therefore, they cannot assure their own clients their AI systems are secure because they can’t actually be assured of vendor security. 
Three of the fifteen participants commented that existing security measures can serve technical robustness in at least some part of their AI. One participant noted that they have five layers to make sure their internal models work, so that was not something they are worried about as it’s one of the easiest ones to secure. Another explained that they have good plans in place for a long time for high availability systems, including backup buildings and risk mitigation processes. A third participant said they hadn’t worked in many systems where there were a lot of concerns around cyberattacks. Even in their current role where there’s a lot of sensitive data, they felt there’s not much cyber risk. Although there was an increased interest in security from customers, their model and security systems hadn’t changed. Two of these three participants worked in non-technical roles, and one worked as a cybersecurity analyst.  

\paragraph{Literature findings:} The literature reported concerns about evolving threats and the pace at which security is keeping up. The area of AI-specific security controls to address evolving threats is a very young but active field of research (\cite{hamon2020robustness}). This paper also found that AI systems being developed did not come close to meeting the minimal requirements of safety and security that would be expected from autonomous systems. 
Evasion attacks, which are attacks on AI models during the testing phase, are the most prevalent type of attacks on machine learning models \cite{oseni2021security}. Poisoning attacks, which are attacks during the training phase, are less common but can easily be carried out on applications that use data from untrusted sources to train their models. The researchers present a systematic framework for demonstrating AI attack techniques. However, they explain that unknown unknowns pose a significant risk as the field of AI evolves, and further research is needed in this area to help identify these. AI systems can produce outputs beyond human comprehension, making it impossible to predict the possibilities in terms of safety risks (\cite{yampolskiy2024monitorability}). The importance of continued research to develop more secure AI systems was echoed by \cite{hossain2024review}, who found that repercussions of successful adversarial attacks can cause harm to public safety and privacy. In addition to discussing data poisoning, exploration, evasion and membership inference, the paper highlights the risks of model inversion, which has the capability to reverse engineer the sensitive data used to train the model from AI model outputs. The rise in adversarial attacks in AI was also reported that research into this field is extremely urgent, including investigating strengthening defence techniques for each individual AI application \cite{kong2021survey}.

In the area of AI safety governance, researchers have proposed independent audits of AI systems to address assurance challenges based on three "AAA" governance principles: Assessments, Audit Trails, and Adherence to jurisdictional requirements (\cite{falco2021governing}). They also stress the importance of interdisciplinary approaches to governance and propose automated real-time assessments of AI systems to increase transparency and reduce risk. There is a need for semi-automated tools for a comprehensive evaluation of AI systems. \cite{mccormack2024ethical}. The cost of developing effective governance systems could be high; however, smaller, more agile sector-based regulators could counter the resource costs associated with the necessary proliferation of regulatory bodies (\cite{falco2021governing}). The approach to governing AI at an industry sector level was echoed by other researchers \cite{mccormack2024comprehensive}, along with calls for the development of context-specific standards \cite{lee2019context}; \cite{de2021artificial}; \cite{ojewale2024towards}. Researchers also reported that countries such as South Korea have already issued Trustworthy AI Development Guidebooks at a sector level (\cite{park2024tough}. Researchers have pointed to existing audit approaches, such as the audit model used in financial accounting, which is based on the Generally Accepted Accounting Principles (GAAP), safety standards such as ISO-13489:2015 and BS8611-2016, which includes a risk assessment for the ethical design of robotic systems \cite{falco2021governing}. 

\paragraph{Discussion:} The field of security for AI systems is not keeping pace with the advances in AI technology, with calls for a collaborative, interdisciplinary approach to developing evaluation and risk mitigation processes for AI systems. The level of unknowns in this space is seen as a high risk, particularly in areas which could result in significant negative societal outcomes in the case of security failure. The rapid pace of technological advances has led to evolving threats such as malicious attacks, data leakage, and increased cyber-attacks. The volume increase is in part due to the ability of AI-powered attacks to move quickly, along with a lower barrier to entry for hackers who can launch cyber-attacks using AI without specialist coding knowledge. Security assessments such as ISO27001 are not seen as fit for purpose for AI systems. Both the interviews and literature question the suitability of point-in-time checklist-based certification, noting that it is too manual and slow to implement and update to be considered a fully comprehensive solution for AI security evaluation. There is a clearly established need for AI technology to be evaluated on an ongoing basis through independent audits, context-specific standards, and real-time automated assessments to enhance transparency and mitigate evolving risks. 

\subsection{Environmental and Societal Impact of AI}
This section explores the wider impact of AI on the environment and society, highlighting concerns about energy consumption, environmental costs, societal inequalities, political radicalisation, and the growing need for ethical governance and cultural sensitivity in AI development.

\begin{table*}[htbp]   
\centering   
\caption{Detailed Findings – Environmental and Societal Impact of AI} 
    \begin{tabular}{|p{5cm}|p{10cm}|}
      \hline
      \textbf{Category} & \textbf{Details} \\
      \hline
      Environmental Impact &
      \textbf{Challenges:}
      \begin{itemize}
        \item High energy consumption of data centres [P1, P2, P4, P6, P9, P11, P13]
        \item Need for fines and regulation on environmental impact of AI [P1, P2]
        \item Concerns About Data Redundancy and Management [P1, P3]
        \item Lack of public awareness \& disconnect from environmental impact of AI [P6, P9]
      \end{itemize}
      \textbf{Observations:}
      \begin{itemize}
        \item Variance in views on environmental benefits vs environmental costs [P3, P4, P9]
      \end{itemize} \\
      \hline
      Societal and Cultural Impact &
      \textbf{Challenges:}
      \begin{itemize}
        \item Positive changes to western workforce, with negative impacts in third-world countries [P7, P9, P10]
        \item Significant negative influence on society and political radicalisation [P2, P3, P9, P14]
        \item Unwanted reinforcement of social inequalities and unrealistic standards [P11, P12, P13]
        \item Loss of culture [P8, P14]
        \item Challenges preventing companies implementing AI with negative societal outcomes [P9, P12, P14]
      \end{itemize} \\
      \hline
    \end{tabular} 
\end{table*}

\subsubsection{Environmental Impact}
This section explores the environmental implications of AI, focusing on resource consumption, sustainability considerations, and the broader impact of AI infrastructure. Participants voiced concerns about the high energy demands of AI, inefficient data storage practices, and the lack of transparency regarding AI’s environmental footprint. Below is a detailed summary of the interview findings, subsequent literature review, and a discussion on both.

\paragraph{Interview findings:} Seven participants expressed concerns about the high energy consumption of AI data centres and its impact on the environment. Participants worked with AI and were aware of and alarmed by the high energy demands of these facilities, which were described as rivalling or surpassing the consumption of entire countries. Additionally, one participant highlighted major concerns over the extraction of rare metals necessary for GPUs. 
Two participants called for regulatory measures to mitigate environmental impacts. Suggestions included implementing fines based on energy consumption and enforcing stricter controls to prevent the unnecessary expansion of data centres driven by redundant data generation. One participant gave the example of considering how many pointless screenshots we each have on our phones and pointed out that AI could be generating and storing similarly redundant data at scale without regulations. The inefficiency in data management was also a point of concern for another participant who saw good use cases for AI to help the environment but did not know how to mitigate data management concerns.
Two participants highlighted a lack of public awareness and a disconnect between AI usage and the environmental impact. It was pointed out that most people don’t associate asking a computer to do something with its impact on the environment. They believed that transparency around the environmental impact of flying gave people choices regarding their travel choices but that the same environmental impact transparency did not exist for AI usage. Another participant pointed out that data centres use more electricity in Ireland than private homes, and with the huge traction AI has gained, this is a real concern.
Three participants commented on the benefits of AI versus its impact on the environment. Two participants noted the potential to use AI for the good of the environment, even though it was not being done now. One participant said that none of the main current use cases being prioritised is about helping the environment, but instead, AI is destroying the environment and polluting for sometimes trivial results. The third participant commented that the increased power usage did not concern them currently, as the benefits outweighed their environmental concerns.
Literature findings: The literature paints the impact of AI on the environment in different ways. On one side, researchers have cited AI as a solution to create efficiencies in companies that will lower their carbon emissions, but another view reports significant concerns about the impacts of AI systems on the environment. Findings indicate that AI can lower emissions through efficient processes developed through innovation and reduction in resource-intensive labour practices, in particular in big cities, where larger, older, and non-technology-based industries (\cite{shang2024impact}). AI has the potential to address environmental issues and climate change \cite{nishant2020artificial}. However, those researchers noted that historical precedents cautioned that new technologies could lead to negative unforeseen issues for sustainability when risks are not managed properly. There is a reported threshold effect, showing that AI reduces carbon emissions only after reaching a particular level of deployment, showing that in parts of China, this threshold was not reached \cite{wang2024role}. They also noted that the effect is notable only in certain industries, particularly labour-intensive ones. Researchers reported that accuracy was prioritised over efficiency, meaning that AI developers were not prioritising carbon emissions when developing models \cite{spelda2020future}. They proposed developing multi-objective models, which considered training models to consider both the environmental impact and accuracy. Development of standards and trade-offs between accuracy and multiple trustworthy AI principles, including the environmental impact of AI systems, was also proposed by (\cite{mccormack2024comprehensive}. Accountability for trade-offs in AI systems requires an increased level of oversight and accountability for businesses that can have conflicting agendas when it comes to profitability and ethics \cite{percy2021accountability}; \cite{diaz2023connecting}. Training a single LLM can emit 300,000kg of CO2, the equivalent of 125 round-trip flights between New York and Beijing\cite{Dhar2020}. They also highlighted the focus and prioritisation of companies on model accuracy over energy efficiency, which results in higher emissions, and highlighted a need to balance model performance with environmental impact. The paper also references the negative environmental impact of extracting metals from the earth. 

\paragraph{Discussion:} The relationship between AI and environmental sustainability is intricate. While there are many potential environmental benefits argued in the literature, the current trajectory of balancing environmental gain with the environmental cost is not positive. The literature shows companies prioritise model accuracy over efficiency, leading to increased carbon emissions. The conflict of interest between prioritising business goals with ethical goals was highlighted, along with concerns about allowing organisations to choose their own trade-offs between environmental impact and business goals like accuracy. Questions around the environmental cost of training and operating AI systems, along with environmental considerations such as the extraction of metals for AI hardware, were highlighted by both the industry participants and researchers. An additional concern was highlighted around the environmental cost of the generation of vast amounts of redundant new data being stored in data centres. The disconnect and misconceptions between AI and its negative carbon footprint raised by participants were also seen in the confusion and conflicts in the literature, which, on the one hand, outlined the potential positive benefits of AI for the environment and, on the other hand, drew attention to the concerns over the carbon emissions from AI systems. The lack of clarity and standards to measure the impact of AI systems on the environment was an additional key concern in the literature.

\subsubsection{Societal and Cultural Impact}
This section examines how AI affects society and culture, including its influence on employment, social structures, and ethical considerations in its deployment. Interviewees raised concerns about AI-driven job displacement, biases reinforcing social inequalities, and AI’s role in shaping political discourse and cultural identity. Below is a detailed summary of the interview findings, subsequent literature review, and a discussion on both.

\paragraph{Interview findings:} Two participants pointed out that AI is about enhancing people’s jobs, not replacing them, and that AI can give more meaningful jobs to people. One participant said they were implementing AI cautiously in their organisation, with a focus on ensuring employees felt secure and were aware their jobs were not being replaced. They explained that they were delivering the rollout by being sensitive to the feelings of employees who might have worked at a task for twenty years that was now being automated with AI.  Another participant suggested taxing AI models in cases where there were significant job displacements. While those comments around changes to the workforce were positive, one participant highlighted the negative impact on those working in third-world countries where dangerous jobs were involved in the mining of rare metals. They described the work done by artisanal miners who mine metals for AI systems with their bare hands in the global south as devastating.
Concerns were raised by four participants around the significant influence AI is having on society, with all four drawing attention to AI’s ability to influence politics. The existing harm to society and democracy was noted as already being extremely damaging, with further concerns surrounding the escalation of AI. Two participants raised significant concerns about the use of AI to profile users and feed them tailored versions of news content. This was described as leading to radicalisation and the formation of online groups that both believe are right because of how AI has tailored their version of news content. Participants referenced platforms such as X (formerly Twitter), TikTok, and Reddit as places where people are being delivered harmful information that has negative societal effects. Participants described how these platforms influence you through AI personalisation, with one describing the information as being fed through social media, resulting in political radicalisation. One participant explained that AI can influence you very quickly because it has built a profile on you and knows you. The creation and dissemination of deep fake videos and images online were noted by one participant as having already damaged trust in the news online and undermining democracy, with another referencing that in social settings on trips to the US, they found it is now considered improper social etiquette to discuss politics due to tensions and deep political divides in their society. 

This research also highlighted concerns around unwanted reinforcement of things such as social inequalities and unrealistic standards. One participant noted that vulnerable groups, such as young people, individuals from underprivileged communities, and people of colour, need more transparency so that AI doesn’t cement existing inequalities in society. Another participant gave the example of harmful AI filters that are designed to alter body types or enhance beauty, asking about the effects that it has on that child when the filter gets turned off, and they immediately compare themselves to the AI version. They pointed out that social media companies should be investing more in the mental health effects of their platforms and the tools their platforms are having on the young people who use them. They argued that platforms need to work more on improving features on their platforms that contribute to increased rates of depression and suicide in young people. Another participant provided the example of bias in insurance models, asking who gets to decide how much weight your gender or postcode should have on your premium, explaining that bias does exist for things like this. They explained that models know that men are more likely to crash and bias against them, but when you reduce that bias in the model, it gets traded off with accuracy. They posed the question, “Who gets to decide what’s fair and unfair?” and called for more transparency for people in society into how these models work.  
Two participants flagged loss of cultural identity, beliefs, language and knowledge as a concern. One participant pointed out that there are variances in cultural beliefs and agreements about what is fair and what is discrimination. They explained that when you ask an AI for an answer, people from different parts of the world would have a different right and wrong answer, giving the example of European data holding European beliefs, which wouldn’t likely be accepted in other places such as Afghanistan, which has its own set of beliefs. Another participant detailed how the language used for data going into and coming out of AI systems was becoming standardised, resulting in a loss of local language, colloquialisms, and even industry terminology that was leading to the sacrificing of cultural identities. They described a situation where a global data team wanted to rename local placenames so they could be more easily understood by an AI system. As they foresaw the downstream cultural effects, they decided to fight for their sense of Irish identity and argued against the change, noting, however, that these things aren’t always considered and that there were already many changes to language they had seen in the short time that AI has been used. 
Three participants raised the challenge of enforcing governance on companies that implement AI with negative societal outcomes. One participant explicitly stated that we cannot control how data gets produced and used, stating that it was definitely a concern for them. They pointed out that they are given sole responsibility for pricing for a large number of customers, and although nobody enforces fairness, it was important for them to have integrity towards the model, so it is fair and works for everyone. However, they said that in their experience, they have to work towards a target revenue, and if it’s not matching, they have to bridge the gap; this involves changes in feature selection and the bias that goes into it. They said that although they would like the model to have a lower level of bias, their company requires them to present multiple options of trade-offs between bias and accuracy so that executives can decide the trade-offs, describing the process of allowing machines to implement bias in these decisions as unfair. Another participant explained that if humans making decisions around the AI process are financially incentivised to choose profit, then they will, and decisions around profit and fairness cannot be made by a person with a financial interest in the outcome. They explained agency is more than just consent; humans should have control over the decisions and the outcomes of the AI system, highlighting that the AI should be for humanity's good rather than the company's profit. Challenges regulating AI internationally were also flagged by two participants, with one calling for international cooperation and transparency around which data is moved between regions. Another participant described how AI gives companies an edge, so even if regulation is introduced, the countries that do permit unregulated AI will gain advantages, in particular, countries that are already heavily sanctioned for other activities. 

\paragraph{Literature findings:} Amerish et al 2020.\cite{amershi2020culture} defines culture as the ethical, sociological, technological, and ideological features of a society or social group which define their way of life, work and communication. They suggest that culture is a precedent to the process of knowledge creation and thus influences the perception and construction of knowledge systems. They believe that the emergence of AI systems, which have been shaped by the cultural parameters of the “West” and the current social and economic power structures, is fundamentally shifting the entire social, political and ethical fabric and disrupting the natural environment. They explained that the foundation of modern technological development in the West originates from British social values, namely the utilitarian principle, which seeks to achieve the greatest good for the greatest number and a desire to control the environment for the betterment of human life. In contrast, countries like Japan inherently value group solidarity, social harmony and a reduction of internal conflict, and China holds the traditional cultural values of Confucianism, which stresses the importance of “formal learning processes” and administrative guidance (by the bureaucratic Mandarin class), and Taoism which emphasises living in harmony with the natural order of the universe. The paper suggests that a fresh perspective on which trajectory the technological process should take is needed, suggesting that the Eastern values of interdependence and universal harmony with nature should be revived and integrated into how we shape the future of technology. Other research argues that variances in cultural values need to be considered when developing evaluation criteria for Trustworthy AI technology \cite{mccormack2024comprehensive}. They also highlight a disconnect between what policy makers, AI experts, and a standard non-expert user considers fair, which, along with these differences in culture, show a need for the inclusion of a wide variety of stakeholders to establish norms for deciding what Trustworthy AI looks like. Research suggest that AI is an opportunity to implement sociotechnical change that can help bring about a better, fairer world \cite{joyce2021toward}. They note that while sociologists are partially contributing to the design of AI systems, they need to be involved more to help design technology that is more beneficial for society. A challenge they present is decision-making protocols that favour corporate elites, naming Silicon Valley as a part of the problem leading to the current class, gender and racial inequalities in societies. Silicon Valley and corporate interests have influenced the direction of academic research, noting as a shift towards entrepreneurship in education aimed at fostering new ventures \cite{pique2020role}. Additional concerns about conflicts over the ownership of art and a loss of culture were raised by researchers \cite{kose_2018}. 

Additional societal concerns around the impact of AI on radicalisation were raised by researchers. Researchers reported that AI systems have increased the radicalisation and divide between viewpoints linked to political violence \cite{pique2020role}. The paper highlights the research gap in the harmful political and social effects of AI. Although AI can be used for cyber security purposes, they suggest that this term should be broadened to include the harmful effects of social media and AI systems on societies.  Persuasion into radicalisation is a process involving the weaponisation of words to convince a target audience of a particular narrative \cite{bamsey2023role}. This paper also discuss the use of counter-narratives to combat these, in particular the ones most likely to lead to violence. They explained how AI was used to distribute “fake news” on social media while also being used to help identify and censor terrorist material. They highlighted that the volume of harmful content online means that it would be impossible to remove it all or attempt to rebut it with counter-narratives. However, they highlight that countries should attempt to decrease the credibility gap, address the root causes, and use both online and offline approaches to solve the issues of online radicalisation that lead to violent extremism. There is a need for Extremism, Radicalisation and Hate speech (ESH) detection tools to combat the societal risks associated with social media’s ability to mobilise extremist communities \cite{govers2023down}. To contribute to protecting both free speech and user safety, the paper proposed an ERH context-mining framework which involved ideological isomorphism (radicalisation), morphological mapping (extremism), and outward dissemination (politicised hate speech). They highlighted the need for more research into resolving biases in dataset collection, annotation and algorithmic approaches in this field. 

Researchers reported a dual positive and negative effect on the workforce. While it was noted that AI could replace lower-skilled positions, resulting in improved skill level of the workforce \cite{shang2024impact}, researchers also raised concerns about the current underpaid workforce that is keeping AI systems going (\cite{Dhar2020};\cite{kose_2018}. Challenges in implementing sufficient oversight and accountabilities for companies who can release AI systems that have harmful societal outcomes were also highlighted by several researchers who demonstrated a clear need for standards and proper accountability for organisations \cite{jobin2019global}; \cite{diaz2023connecting}; \cite{percy2021accountability}.

\paragraph{Discussion:} While there are many arguments around the potential of AI for good, the current economic and social structures have led to the prioritisation and shaping of AI technology that is causing unintended harm to societies. Western cultural values and influential decision-makers in places like Silicon Valley have shaped the development of AI technology. However, the decisions made by those investing in technology conflict with the values held by populations in Western societies, which broadly strive for the greatest good for the greatest number of people. The societal values of controlling and subjugating the natural world for the benefit of humans have spread rapidly throughout Western culture in recent years from Great Britain, directly conflicting with the dominant Eastern values of harmony with the natural world. AI systems, the design and function of which are shaped by the cultural, social and economic structures existing in the global technology industries, can transform societies at scale and result in mass shifts in the fabric of our societies. Adverse societal outcomes are seen from AI systems, in particular, the radicalisation of viewpoints online, the amassing of extremist communities and a rise in political violence. AI is expected to cause significant shifts in the global workforce. After a transition period of job displacement, the Western world anticipates positive outcomes: job functions are expected to become more attractive for employees, with a reduction in unwanted, lower-paid administrative tasks. However, there is already a notable reliance on an underpaid workforce, described by researchers as a "sweatshop of programmers" in the Eastern world, which sustains AI infrastructure globally. Further research is needed to establish how we can challenge existing social and economic structures to influence the design of AI systems, which can be aligned with the culture and values held by societies using that technology. 

\section{Reflection on the EU AI Act from our Findings}
The EU AI Act \cite{eu_ai_act_2024} seeks to protect fundamental rights by regulating AI systems, mandating risk assessments, accountability mechanisms, and compliance requirements. It also refers to the EU Principles of Trustworthy AI as a guidance for safe AI development. However, the findings from this study highlight significant gaps in how the industry professionals in this study understand and implement trustworthy AI principles, raising questions about whether the Act’s provisions will be effective in practice. The section links key industry concerns identified in our research to the regulatory approach taken by the AI Act.

\subsection{Accountability, Governance, and Regulatory Compliance:} 
The AI Act, clearly defines the roles and responsibilities for those involved in AI systems, but accountability in practice remains complex. AI accountability is divided across multiple roles in the Act, including AI providers who build, buy or adapt AI models and place them on the market, and deployers who use them. Additional responsibilities fall on importers, distributors, and supervisory authorities, but ambiguity arises when any distributor, importer, deployer or other third party makes a ‘substantial modification’ to a high-risk AI system, shifting their classification and compliance requirements. The Act mandates technical documentation and data labelling transparency, yet challenges such as communication around potentially sensitive data between various stakeholders in the AI development process remain. This includes difficulties coordinating compliance across multiple stakeholders when dealing with third-party AI models (\cite{golpayegani2024ai}. 
Findings from industry professionals during our interviews highlighted key concerns with compliance readiness. Accountability within organisations is fragmented, with unclear ownership of AI governance and reliance on third-party compliance claims that often lack verification. Many companies approach AI governance reactively, addressing risks only when problems arise, while existing compliance tools are insufficient for AI-specific risks. Financial incentives also influence compliance, with auditors and internal teams sometimes prioritising approvals over rigorous assessments. Unlike GDPR, AI regulations lack widely adopted compliance frameworks, making enforcement inconsistent and challenging. 
While the AI Act establishes an important foundation for AI governance, the insights reported by participants in this study and subsequent literature review indicate that its effectiveness depends on stronger enforcement mechanisms. Real-time compliance monitoring, sector-specific regulatory frameworks, and independent auditing are necessary to prevent the Act from becoming another bureaucratic hurdle rather than a meaningful tool for ensuring Trustworthy AI. 

\subsection{Data Management and Quality in AI Systems:} 
The AI Act includes provisions for data governance, requiring that AI models be trained on high-quality, well-documented datasets. However, our interview findings show that poor data quality, provenance issues, and synthetic data manipulation are widespread industry concerns. Many AI professionals noted that their organisations struggle to ensure data integrity, with some questioning whether data brokers provide truthful representations of consent and compliance. The lack of transparency in third-party data sourcing is a direct challenge to the AI Act’s goal of ensuring fairness and accountability. Additionally, the growing reliance on synthetic data raises concerns about bias amplification and the loss of real-world validity in AI models. While the AI Act touches on data quality, it does not yet include specific provisions to regulate synthetic data generation, which may become a loophole that weakens compliance efforts. Given the risks associated with data quality failures, including biased decision-making and unreliable AI outputs, the Act may need to introduce stricter requirements for provenance tracking and independent data audits.

\subsection{Human Factors in AI Development and Oversight:}
The AI Act mandates human oversight in high-risk AI systems, requiring human intervention in decision-making processes. However, industry professionals highlight several challenges with how human oversight is implemented in practice. Some noted that organisational structures do not currently support effective AI governance, as compliance teams often lack the technical expertise to assess AI risks, and AI engineers may not fully understand ethical and regulatory requirements. Another key issue is conflicts of interest in human decision-making. Several professionals pointed out that when humans are included in AI oversight, their incentives may prioritise business interests over ethical concerns. For example, when AI bias is detected, professionals reported that leadership often dismisses these concerns due to the financial trade-offs associated with bias mitigation. The AI Act does not currently specify the qualifications, independence, or ethical responsibilities of human overseers, leaving significant room for organisations to self-regulate oversight in ways that may not be effective. Additionally, there is an emerging risk of over-reliance on AI, where professionals assume that AI-driven compliance tools will inherently prevent unethical or illegal behaviour. This “blind trust” in AI automation contradicts the AI Act’s aim to ensure meaningful human control, highlighting a gap between regulatory intent and industry practice.

\subsection{Technological Robustness and Safety:}
While the AI Act introduces requirements for technological robustness and security, our findings reveal that AI security risks are evolving faster than regulatory frameworks. Industry professionals expressed concerns about data poisoning, adversarial attacks, and sensitive data leakage in AI systems, but noted that security compliance efforts often rely on reactive measures rather than proactive risk management. A number of participants specifically criticised the use of static certifications like ISO27001 for AI security, arguing that these frameworks were not designed for real-time AI monitoring. The Act’s provisions on security could be strengthened by incorporating continuous evaluation mechanisms rather than relying solely on documentation and pre-deployment risk assessments. Without this, companies may remain vulnerable to dynamic security threats that emerge after AI systems are deployed.

\subsubsection{Environmental and Societal Impact:}
The AI Act includes fairness and societal well-being as core principles but does not yet introduce concrete enforcement mechanisms for evaluating AI’s societal and environmental impact. Our research finds that AI professionals are increasingly aware of AI’s negative societal consequences, including reinforcement of social inequalities, political radicalisation, and cultural loss. However, companies currently lack clear guidelines on how to measure and mitigate these risks in practice. Similarly, AI’s high energy consumption remains an industry concern, with some professionals calling for environmental impact assessments and potential fines for AI-driven carbon footprints. While the AI Act does acknowledge the need for responsible AI development, it does not yet introduce binding sustainability requirements. As AI adoption continues to grow, the regulatory framework may need to evolve to include mandatory environmental impact assessments for AI models, particularly those requiring large-scale computational resources.

\section{Practical Checklist for Operationalisation of Trustworthy AI}\label{sec6}
While this study has identified critical challenges in achieving trustworthy AI across governance, data quality, regulatory compliance, fairness, and human oversight, it is evident that the industry lacks a consolidated, actionable pathway for addressing these issues in practice. Many of the principles outlined in the EU's Ethical Guidelines for Trustworthy AI, though widely accepted in theory, remain difficult to implement. To bridge this gap between theory and application, we propose a practical governance checklist with a list of actions designed to assist organisations in operationalizing the challenges identified in this paper for the implementation of Trustworthy AI. This is detailed in table \ref{table8} Although standards such as ISO42001\cite{ISO42001} offer structured frameworks for AI Management Systems, they are often high level and lack the focus of the specific challenges which were identified through this research. This checklist translates the thematic findings of our interviews into concrete organisational actions that can be implemented across organisations, as a supplement to existing certifications and standards such as ISO42001. While this checklist is designed for practitioner use, we acknowledge that its application may vary considerably across sectors and organisations and we therefore frame it as a flexible guide rather than a prescriptive standard.

\begin{table*}[htbp]
  \centering
  \caption{Key Challenges and Recommended Organisational Actions}
  \label{table8}
  \begin{tabular}{|p{6cm}|p{10cm}|}
    \hline
    \textbf{Key Challenge} & \textbf{Recommended Organisational Action} \\
    \hline
    Siloed departments \& communication gaps & Establish cross-functional AI steering committees to improve collaboration and compliance integration. \\
    \hline
    Unclear lines of responsibility & Clearly define and document roles and responsibilities for AI compliance and accountability. \\
    \hline
    Blind trust in third-party compliance & Conduct thorough due diligence and regular audits of third-party providers. \\
    \hline
    Lack of systematic bias testing & Implement regular and mandatory bias audits and fairness evaluations. \\
    \hline
    Conflict between fairness and profitability & Incorporate fairness metrics into business KPIs and decision-making frameworks. \\
    \hline
    Confusion around AI Act implications & Provide targeted training on regulatory requirements and implications for different teams. \\
    \hline
    Lack of AI-specific standards & Adopt and contribute to the development of sector-specific AI standards. \\
    \hline
    Rush to implement AI without risk mitigation & Mandate risk assessments before deployment of AI systems. \\
    \hline
    Lack of AI assessment tools & Invest in or develop AI-specific compliance and assessment tools. \\
    \hline
    Poor data quality & Introduce automated data quality checks and establish quality thresholds. \\
    \hline
    Vendor misrepresentation of data consent & Implement stringent data provenance checks and vendor contracts with accountability clauses. \\
    \hline
    Overreliance on synthetic data & Use synthetic data cautiously, ensuring validation and bias monitoring processes. \\
    \hline
    Inadequate data documentation & Develop automated and regularly updated data lineage documentation systems. \\
    \hline
    Insufficient human oversight & Define human-in-the-loop checkpoints and mandate oversight at critical decision points. \\
    \hline
    Lack of explainability & Use explainable AI methods and ensure model transparency is available to internal and external stakeholders. \\
    \hline
    Model hallucinations & Ensure high-quality, up-to-date data is used and monitor model outputs for anomalies. \\
    \hline
    Evolving cyber threats & Establish continuous AI security assessment protocols and update cybersecurity training. \\
    \hline
    Increased surface area for attacks & Limit unnecessary data exposure and API access points through stringent access controls. \\
    \hline
    Unethical data practices & Enforce ethical data collection policies and increase internal auditing of data usage. \\
    \hline
  \end{tabular}
\end{table*}

\section{Discussion and Future Directions}\label{sec7}
AI technologies present both opportunities and challenges in ensuring the development and deployment of TAI. Addressing the identified industry challenges requires an independent and sector level approach involving policy reforms, standardisation of industry best practices, and collaborative efforts including all relevant stakeholders. This section outlines future directions for the improvement and adoption of TAI principles including recommendations for policymakers, industry and academia. All three areas should involve diverse stakeholders, including underrepresented communities, in their decision-making process to ensure AI systems are developed with a broader perspective, mitigating biases and promoting fairness across different segments of society. Efforts should be made to prevent the loss of cultural identities due to AI standardisation. This can be achieved by incorporating local languages, customs, and values into not just developing AI systems, but at the level of prioritising which technology gets researched, developed and implemented into societies. The use of AI in social media should have sufficient oversight to prevent the spread of misinformation and radicalisation, whilst also respecting and preserving cultural diversity of the societies they introduce their systems into. Implementing AI models that detect and mitigate harmful content can help protect societal well-being, but due to the speed of which AI can be rolled out, the cultural impact should also be prioritised.

\subsection{Policymakers}
This study’s findings indicate that the AI Act alone may not fully address the practical challenges of AI implementation. Industry professionals face uncertainty regarding compliance, struggle with data quality and oversight, and report that security risks are escalating beyond what current regulations can manage. Without stronger enforcement mechanisms, sector-specific governance frameworks, and real-time AI evaluation tools, the Act may prove insufficient in ensuring genuinely Trustworthy AI. To close the gap between policy and practice, regulators should clarify accountability structures to ensure independent and enforceable AI oversight roles, introduce dynamic compliance mechanisms that go beyond static certifications to incorporate real-time AI system monitoring, and enhance security provisions by requiring ongoing adversarial testing and AI-specific risk mitigation measures. Additionally, regulating synthetic data usage through stricter standards for provenance and bias mitigation, as well as incorporating environmental accountability by mandating carbon footprint assessments for AI models, could strengthen the Act’s effectiveness. By addressing these gaps, the AI Act could evolve from a broad regulatory framework into a more effective instrument for ensuring AI accountability, security, and ethical alignment in real-world applications.

Policymakers should work towards providing clearer and more detailed regulatory guidelines at an industry sector level to eliminate ambiguities surrounding the EU AI Act and other related regulations. The creation of AI-specific standards and certifications, audited by independent bodies who are not being paid by the company they are auditing, is crucial for establishing uniform and fair evaluation. Standard-setting organisations, in collaboration with industry experts and academics, should develop quantifiable metrics and real-time assessment tools tailored to different AI use cases and sectors. These standards should address all seven TAI principles. To enhance accountability, regulators could introduce independent auditing bodies to oversee AI compliance at a sector level. These bodies should operate without conflicts of interest to ensure impartial evaluations. In addition to developing quantifiable standards, enforcing stricter penalties for non-compliance, such as higher fines and restrictions on operations is required to ensure that companies are sufficiently deterred from bypassing ethical considerations in favour of profitability. Given the global nature of AI development, international cooperation is essential for harmonising regulations and standards. Policymakers should engage in cross-border dialogues to address challenges related ensuring that AI systems are developed safely, and in alignment with the values of the societies that they are impacting.

\subsection{Industry} 
In addition to the recommended actions provided in\ table \ref{table8}, organisations should embed ethical principles into every stage of the AI development lifecycle. This includes conducting thorough bias testing, ensuring data quality, and incorporating fairness and transparency metrics. Companies can build more trustworthy AI systems by prioritising ethical considerations alongside technical performance. Forming interdisciplinary teams or AI steering committees can break down departmental silos and foster collaboration between technical experts, compliance officers, and business leaders. These teams should oversee AI strategy, compliance, and risk management, ensuring that accountability is clearly defined within the organisation. Continuous education and training programmes can enhance employees' understanding of AI ethics, data literacy, and regulatory requirements. By equipping staff with the necessary knowledge and skills, organisations can improve human oversight and reduce risks associated with over-automation and poor decision-making. Transparency in AI systems can build trust among stakeholders. Organisations should adopt tools and methodologies that explain AI decision-making processes, making them accessible to both internal teams and external users. Transparent communication about data sources, model decision making and limitations, and ethical considerations can also mitigate misunderstandings and increase user trust and adoption of AI systems.

\subsection{Academia}
Academic institutions and research organisations should encourage multidisciplinary studies that combine insights from areas such as computer science, ethics, sociology, and law. These collaborations can lead to the development of more holistic approaches to TAI design and evaluation and help to address complex challenges like cultural biases and societal impacts. There is a pressing need for semi-automated and real-time assessment tools capable of evaluating AI systems comprehensively. These tools should align with regulations such as the AI act, and include a number of metrics for TAI principles, which can be adopted for multiple use cases. Research efforts should focus on creating solutions that can continuously monitor AI performance and help inform decisions around trade-offs for TAI principles and company goals such as profitability. This will increase transparency and aid in reducing the reliance on manual checklists and point-in-time assessments. Educational initiatives aimed at the general public can bridge the disconnect between AI usage and its societal and environmental impacts, as well as increase knowledge about how these systems work and their biases and limitations. By increasing awareness, individuals can make informed decisions and advocate for responsible AI practices, contributing to a culture that values TAI principles. Collaborative efforts should be directed towards researching energy-efficient AI models and promoting sustainable practices. This includes optimising algorithms for lower energy consumption, utilising renewable energy sources for data centres, and developing standards to measure and report the environmental impact of AI systems. Thought should be put into what aspects of AI academia wants to help further knowledge in. Current economic structures, including corporate interests, should not continue to increase their influence in directing which aspects of knowledge is furthered by academic research. 

\subsection{Limitations}
The study has several limitations. First, the sample was confined to the UK and Ireland, where regulatory proximity to the EU AI Act may shape perspectives in ways that differ from those in other jurisdictions. Second, while we sought diversity in job functions, certain roles central to AI development (e.g., algorithm engineers, UX researchers, in-house legal experts) were not included. Their absence limits the generalisability of our findings, though the focus on deployment and governance roles remains aligned with our research objectives. Finally, with fifteen participants, our qualitative approach prioritised thematic depth over breadth; as such, findings should be interpreted as illustrative of lived industry perspectives rather than definitive of the sector as a whole.

\subsection{Proposed Next Steps}
While this paper prioritises empirical insight from industry voices, we recognise that further work could benefit from deeper integration with established theoretical frameworks such as institutional theory, organisational ethics, or sociotechnical systems thinking. These lenses offer valuable ways of understanding how organisational norms, structures, and power dynamics shape the adoption of Trustworthy AI. However, given the emergent and still-contested nature of Trustworthy AI as a concept, and the intention of this paper to centre lived experience and practical barriers, we have chosen not to impose a singular theoretical framing. Future research could productively explore how these perspectives might complement the findings presented here, particularly in understanding how ethical AI commitments are shaped, constrained, or enabled by institutional context. 

As a next step, we are focusing on developing a follow-up project that builds on the findings of this paper to provide more practical guidance for industry. This work will focus on designing a governance evaluation framework to help organisations align with both the Trustworthy AI principles and the requirements of the EU AI Act. It will translate high-level ethical and regulatory expectations into a practical tool that can be used across different stages of the AI lifecycle to assess systems for both regulatory compliance and adherence to the seven Trustworthy AI principles. Informed by the challenges identified through our interviews, the proposed framework will aim to bridge the gap between abstract principles and day-to-day industry implementation. Future work could more explicitly integrate theoretical frameworks, such as institutional theory, to situate these findings within broader organisational research traditions. In this study, however, our priority was to focus on the industry participants’ lived experiences in practice.

\section{Conclusion}
This paper contributes significantly to academia by providing empirical insights into the challenges faced by industry professionals who participated in this study in implementing Trustworthy AI. Through structured interviews with fifteen experts across various sectors, we bridge the gap between academic and regulatory knowledge and real-world TAI practices of the participants interviewed. The study highlights ambiguities in accountability and governance, revealing a "wild west" mentality where both AI adoption and enforceable regulations are not incorporating thorough ethical considerations. By documenting issues raised by participants such as data quality problems, inherent biases, and overreliance on AI without proper safeguards, the research emphasises the urgent need for standardised metrics, automated assessment tools, and interdisciplinary oversight. Academically, it fills a critical gap by offering qualitative data on industry practices, underscoring discrepancies between policy intentions and industry realities. It also highlights a need for societal and environmental impacts to be included in safety assessments of AI technology. In essence, this study advances academic knowledge by providing empirical evidence of industry challenges in the development and assessment of TAI, highlighting the need for more tangible regulations and standards, and emphasising the important role of human oversight and accountability. The findings from the participants in this study illustrate how practitioners in our sample perceive and navigate the challenges of Trustworthy AI. While the insights are not generalisable to the industry as a whole, they nonetheless highlight pressing issues in governance, accountability, and evaluation that are likely to resonate across a range of organisational contexts. These insights lay a foundation for future research to develop effective strategies and policies that align AI development with societal values and ethical principles.

\backmatter

\bmhead{Authors and Affiliations}
All authors have reviewed and consented to the publication of the manuscript as presented. This research received partial support from the Research Ireland under grants 13/RC/2106\P2 (ADAPT) and is co-funded by the European Regional Development Fund (ERDF).

\bmhead{Ethics Declaration}
The data employed in this review are sourced from one to one interviews with fifteen professionals following the University of Galway Ethics Process. The research was approved by the University of Galway’s Research Ethics Committee (REC). The ethics protocol covered the outreach approach, which included contacting participants via email and LinkedIn. Informed consent was obtained from all participants via signed consent forms prior to their involvement in the structured interviews. This ethical process ensured compliance with GDPR regulations and ethical standards for research involving human participants.

Additionally research was collected through publicly available materials, including published research articles, ISO standards, books, and openly accessible databases and industry publications. All sources are duly cited and listed in the reference section of this paper. 
\onecolumn
\begin{appendices}

\section{Structured Interview Questions}\label{secA1}
\paragraph{Data Acquisition}
    \begin{itemize}
      \item Do you know how data is acquired for use in AI typically, and if so, can you tell me about it?
      \item In your opinion, what are the most important aspects of data acquisition for AI?
      \item Do you have concerns around data acquisition for AI?
    \end{itemize}
\paragraph{Data Quality}
    \begin{itemize}
      \item What do you know about data quality and its importance in relation to AI?
      \item In your opinion, what are the most important aspects of data quality for AI?
      \item Do you have concerns around data quality for AI?
    \end{itemize}
\paragraph{Data Preparation}
    \begin{itemize}
      \item Do you know how data is prepared for use with AI? Tell me about it if so.
      \item In your opinion, what are the most important aspects of data preparation for AI?
      \item Do you have concerns around data preparation for AI?
    \end{itemize}
\paragraph{Data Provenance}
    \begin{itemize}
      \item Do you know if documentation is generally kept on data as it moves through the various stages of the AI system?
      \item If you wanted to understand aspects of data usage in the company, how would you find out where the data came from and how it was transformed or used?
      \item In your opinion, what are the most important aspects of data provenance for AI?
      \item Do you have concerns around data provenance for AI?
    \end{itemize}

\paragraph{Trustworthy AI}
    Do you have any additional comments, concerns or insights in relation to AI, specifically in relation to each of the seven key principles:
    \begin{enumerate}
      \item Human agency and oversight
      \item Technical robustness and safety
      \item Privacy and data governance
      \item Transparency
      \item Diversity, non-discrimination and fairness
      \item Environmental and societal well-being
      \item Accountability
    \end{enumerate}

    \paragraph{Assessment of Standards and Regulations} 
    \begin{itemize}
      \item How do organisations you have worked with generally assess for compliance with standards like ISO27001 and ISO27701 or regulations like GDPR?
      \item Can you tell me any methods or frameworks you are aware of?
      \item In your experience have you seen any challenges with assessment?
      \item In relation to self-assessment for AI compliance in the five aspects of data we previously discussed, can you foresee any potential challenges or obstacles?
      \item How does your organisation document assessment with standards and regulations?
      \item Are there any issues you see in relation to documenting these assessments?
      \item Who are the key stakeholders involved in the assessment of standards and regulations like GDPR and ISO27001?
      \item Do you know what processes these stakeholders follow or how they engage with the wider organisation on this work?
      \item Based on your knowledge of ISO27001 and ISO27701 (if any), what are the key aspects you think will need to be extended to include AI-specific aspects of these ISOs?
    \end{itemize}




\end{appendices}





\bibliography{main}


\end{document}